\documentclass[aps,pre,aps,showpacs,singlecolumn,notitlepage]{revtex4-1}
\usepackage[english]{babel}
\usepackage[applemac]{inputenc}
\usepackage{graphicx}
\usepackage{amsmath,bbm}
\usepackage{amsfonts}
\usepackage{amssymb}
\usepackage{booktabs}
\usepackage{epsfig}
\usepackage{psfrag}
\usepackage{pstool}
\usepackage{epstopdf}
\usepackage[caption=false]{subfig}
\usepackage{appendix}
\usepackage{bm}
\usepackage{ulem}
\usepackage{colortbl}
\usepackage{mathrsfs}
\usepackage{cleveref}	
\usepackage{bbold}

\newcommand\figref[1]{Fig. \ref{#1}}

\newcommand\secref[1]{Sec.~\ref{#1}}
\newcommand\appref[1]{Appendix~\ref{#1}}

\DeclareMathOperator\erf{erf}

\newcommand{\ii}{\mathbbm{i}}

\newcommand{\E}{\mathrm{e}}
\newcommand{\D}{\mathrm{d}}
\newcommand{\G}{\mathrm{g}}

\newcommand{\ls}{{\lambda^*}}
\newcommand{\lss}{\lambda^{**}}
\newcommand{\my}{J_Y}
\newcommand{\myss}{J_Y^{**}}

\newcommand{\average}[1]{\left<{#1}\right>}
\newcommand{\p}[1]{\left({#1}\right)}
\newcommand{\pq}[1]{\left[{#1}\right]}

\begin{document}

\title{Efficiency fluctuations in cyclic machines}

\author{Marc \surname{Su\~n\'e}}
\email{msune@phys.au.dk}
\affiliation{
Department of Physics and Astronomy,
Aarhus University,
DK-8000 Aarhus C, Denmark}

\author{Alberto \surname{Imparato}}
\email{imparato@phys.au.dk}
\affiliation{
Department of Physics and Astronomy,
Aarhus University,
DK-8000 Aarhus C, Denmark}

\begin{abstract}
We study the statistics of the efficiency in a class of isothermal cyclic machines with realistic coupling between the internal degrees of freedom. We derive, under fairly general assumptions, the probability distribution function for the efficiency. We find that the macroscopic efficiency is always equal to the most likely efficiency, and it lies in an interval whose boundaries are universal as they only depend on the input and output thermodynamic forces, and not on the details of the machine. The machine achieves the upper boundary of such an interval only in the limit of tight coupling.  Furthermore, we find that the tight coupling limit is a necessary, yet not sufficient, condition for the engine to perform close to the reversible efficiency. 
The reversible efficiency is the least likely regardless of the coupling strength, in agreement with previous studies. By using a large deviation formalism we derive a fluctuation relation for the efficiency which holds for any number of internal degrees of freedom in the system.

\end{abstract}

\date{\today} 
\maketitle

\section{Introduction}
\label{Intro}
Since the dawn of thermodynamics, the capacity of a machine to convert available sources of energy into useful work has been an ubiquitous subject of investigation. As a matter of fact, the second law of thermodynamics sets a limit on a thermal machine's performance.
In particular, the maximum efficiency of an isothermal machine, as given by the ratio of work performed by the machine to the energy used, is 1. The lossless limit in which energy conversion into work is performed with efficiency 1 is nonetheless attained in the reversible quasi-static limit, in which the machine operates infinitely slowly. A machine in this reversible quasi-static regime delivers zero output power, and so it is useless for practical purposes. Accordingly, many efforts have been devoted to the study of the condition for finite non-zero, possibly maximal,  power production. One of the first discussions on this topic is attributed to Moritz von Jacobi already around 1840~\cite{Jacobi1837}.

The blossoming of experimental techniques aimed at investigating the fluctuations of thermodynamic quantities in microscopic systems~\cite{Ciliberto17} has paved the way to the extension of the laws of thermodynamics to address the stochastic properties of quantities such as work, heat or entropy production~\cite{Seifert2012}. According to this revised version of thermodynamics, the efficiency itself is a fluctuating quantity~\cite{Verley2014} as it is  given by the ratio of two fluctuating quantities: the entropy production rates $\sigma_i$ associated  to the output ($i=2$) and input ($i=1$)  currents along a single stochastic trajectory 
\begin{align}
\eta=-\frac{\sigma_2}{\sigma_1}.
\label{eq_eff-Intro}
\end{align}
As such, the trajectory dependent efficiency of a microscopic machine performing at the energy scale of the thermal fluctuations ($k_BT$) \cite{Verley2014} can indeed surpass the reversible limit (or the Carnot limit for thermal motors).
 Furthermore, collective effects such as synchronization in arrays of $N$ interacting microscopic motors can decrease the energy dissipation~\cite{IzumidaKori16} and possibly increase  the thermodynamic efficiency with respect to the single motor case ~\cite{Golubeva2012a,Golubeva2013,Golubeva2014,Imparato15,VroylandtEsposito17}, and even beat the  Carnot limit at  finite entropy production rate~\cite{Campisi2016}. 
The study of the statistical properties of the stochastic efficiency is thus of crucial importance in order to characterize the performance of microscopic machines operating in out-of-equilibrium conditions.


In this paper we derive via stochastic thermodynamics the statistics of the efficiency for a class of cyclic isothermal energy transducers~\cite{Zemansky11,Callen06}, whose internal degrees of freedom are coupled with  realistic physical interactions described by a many--body potential.  Starting from the simplest case of a machine consisting of two degrees of freedom and concluding with  the $N$-particle  system, we are able to derive the full probability density function (PDF) of the efficiency  under fairly general assumptions. The efficiency PDF is known to exhibit power law long tails~\cite{Proesmans15,Polettini15}, and as such finite moments of any order cannot be calculated. 
However, our approach allows us to identify the macroscopic efficiency with the most likely value, i.e. the maximum of the efficiency PDF.
Furthermore, the mechanistic description used here allows us to derive the exact expression of the machine response as a function of the force intensity: this in turn allows us to accurately study the weak and tight coupling limits, and the large input/output force regime, so as our investigation is not limited to the linear regime.

As far as the least likely efficiency is concerned, we find that it corresponds to the reversible efficiency, in accordance with the findings of Ref.~\cite{Verley2014}. In that reference the fluctuation theorem for the entropy production ~\cite{Seifert05a,Imparato06,Seifert2012} was used to prove this result on the least likely efficiency. Here we take one step further, and show that the fluctuation theorem for the energy currents \cite{Imparato7a,Fogedby12,Fogedby14,Imparato2014}  implies a fluctuation relation for the efficiency itself: the PDF of $\eta$ turns out to show a symmetry which resembles those obtained previously for, e.g., the work or the heat PDFs \cite{Seifert2012,Imparato06,Ciliberto2013,Ciliberto2013a,Berut16,Berut16a}.
While we initially assume that the input and output energy currents are Gaussian distributed as, e.g., in~\cite{Polettini15,ProesmansDreher16}, we provide solid evidence that the fluctuation relation for the efficiency holds beyond the linear regime, and for a general interaction potential.

The paper is organized in the following manner. In section \ref{Sec1} we review a few useful results on the Brownian particle in a tilted periodic potential, which will be used in the following discussion in the paper. In section \ref{Sec2} we consider the minimal model for an isothermal cyclic energy transducer, namely a system with two degrees of freedom and a periodic interaction potential. We then derive the efficiency PDF, discuss its extremal points, and introduce the fluctuation relation for $\eta$. In  section \ref{Sec3} we generalize our results to the case of a machine with $N$ degrees of freedom. In section \ref{Conclusions} we summarize our results.

\section{Single oscillator}
\label{Sec1}
A Brownian  particle in a one-dimensional periodic ring potential $U_0(y)$ and driven by a force $f$ is the minimal model for the study of isothermal systems driven into a non-equilibrium steady state~\cite{SpeckSeifert06,BlickleSpeck07,Golubeva2012,VandenBroeck2012,Speck12,NyawoTouchette16}. Furthermore, its properties are relevant for the study of a system with many degrees of freedom, interacting through periodic potentials, as we argue in the following sections. We thus review some of its features and include a few novel results as well in this section. 

The trajectory $y(t)$ of an  overdamped Brownian  particle in a periodic potential $U_0(y)$ with period $L$  and subject to a constant drift force $f$ is generated by the Langevin equation 
\begin{align}
\dot y=f-U_0'(y)+\zeta(t),
\label{eq_Langevin-Sec1}
\end{align}
where the friction coefficient is set to unity $\Gamma=1$, and  a dot and a prime indicate time and space derivatives, respectively. The quantity $\zeta(t)$ is a stochastic force with a Gaussian distribution and correlations given by the fluctuation--dissipation relation
\begin{align}
\langle\zeta(t)\,\zeta(t')\rangle=2 T\delta(t-t'),
\label{eq_white_noise-Sec1}
\end{align}
that accounts for thermal fluctuations due to energy exchange between the system and the surrounding medium at temperature $T$. The Boltzmann constant $k_B$ is set to unity throughout this paper.\\
Furthermore, in the following the quantity $k$ will express the typical amplitude of the periodic potential corrugations, the simplest example being $U_0(y)=-k \cos y$. We will not assume any specific for $U_0(y)$, unless differently stated.

The equation for the time evolution of the probability distribution function (PDF) of the phase $y$ reads
\begin{align}
\partial_tP(y,t)=\mathcal{L}_yP(y,t),
\label{eq_FP_y_Sec1}
\end{align}
where $\mathcal{L}$ is the Fokker--Planck (FP) differential operator
\begin{align}
\mathcal{L}_y=-\partial_y(f-U_0'(y)-T\partial_y).
\label{eq_op_FP_y-Sec1}
\end{align}
The PDF in the steady state is  thus~\cite{Imparato15,vanKampen1981,Risken}
\begin{align}
P(y)=\mathcal{N}\beta\E^{\beta(-U_0(y)+f y)}\left[\frac{I(L)}{1-\exp{(-\beta Lf)}}-I(y)\right],
\label{eq_pdf_ss-Sec1}
\end{align}
where $I(y)=\int_0^y \D y'\exp{[-\beta(-U_0(y')+f y')]}$, $\beta=1/T$, and $\mathcal{N}$ is a normalization constant that depends implicitly on $\beta$, $k$ and $f$,  and which is fixed by the normalization condition
\begin{align}
\int_0^LP(y)\D y=1.
\label{eq_normalization-Sec1}
\end{align}
The steady--state PDF as given by Eq.~(\ref{eq_pdf_ss-Sec1}) has the same periodicity as the potential $U_0(y)$.
The steady--state velocity of the dynamical variable $y$ reads~\cite{Golubeva2012,Imparato15}
\begin{align}
\bar v_y(k,f)=L\,\mathcal{N},
\label{eq_ss_vel-Sec1}
\end{align}
in which the dependency on the temperature is implicit. This is an exact result that holds for any potential strength $k$. The velocity $\bar v_y$ depends in particular on the form of the potential $U_0(y)$. However, the asymptotic behaviors can be predicted by using some physical arguments: a) in the limit of large corrugation amplitude  ($k\gg T,\, f$) the particle is effectively trapped in a potential well, and so $\bar v_y\to 0$; b) in the opposite limit ($k\ll T,\, f$) the potential is flattened by the tilting force, hence $\bar v_y\to f$. We can thus express the steady--state velocity in terms of a function $c(k,f)$,
\begin{align}
\bar v_y(k,f)=f[1-c(k,f)],
\label{eq_ss_vel_c-Sec1}
\end{align}
such that $0\le c(k,f)\le 1$, $c(0,f)=0$ and $c(\infty,f)=1$. Finally,  we notice that the integrals contained in the expression for the normalization constant Eqs.~\eqref{eq_pdf_ss-Sec1}-\eqref{eq_normalization-Sec1} typically do not have an analytic solution, though the steady--state velocity can be expanded in power series of $k$~\cite{Imparato15}.\\

\subsection{Stochastic work}
\label{Sec1-1}
The total work done on the particle along individual trajectory is defined by the functional~\cite{Sekimoto97,Sekimoto98}
\begin{align}
w_y[y(\tau)]=\int_0^t f \dot y(\tau)\,\D\tau=f\cdot(Y_t-Y_0).
\label{eq_work-Sec1-1}
\end{align}
Here we have introduced a second coordinate $Y$ to account for the total traveled distance: such coordinate is unbounded ($-\infty<Y<\infty$) in contrast to the bounded periodic coordinate $y$. The stochastic processes for $y$ and $Y$ (and hence $w_y$) are characterized by the same Langevin equation Eq.~\eqref{eq_Langevin-Sec1}, the only difference being that the former coordinate is periodic while the latter is unbounded. In particular both coordinates have the same velocity in the steady-state $\langle \dot Y\rangle=\average {\dot y}$.

 The coordinate $Y$ represents a time integrated current for the Brownian particle, and the study of its fluctuations is propaedeutic to the subsequent study of the efficiency fluctuations. In particular, we notice that the time evolution of its PDF is governed by the analogous evolution operator to that for the variable $y$ Eq.~\eqref{eq_op_FP_y-Sec1}:  $\partial_t P(Y,t)=\mathcal{L}_Y P(Y,t)$.

\subsection{Fluctuations of $Y$}
\label{Sec1-4}
In view of studying the fluctuations of the variable $Y$, it is convenient to introduce the evolution operator $\mathcal{\hat L}$ for the joint probability $P(y,Y,t)$ that reads~\cite{Imparato07,Imparato08,Mehl08,Fogedby12,Fogedby14}
\begin{align}
\mathcal{\hat L}=-\partial_y(f-U_0'(y))-\partial_Y(f-U_0'(y))+T(\partial_y^2+\partial_Y^2+2\partial_y\partial_Y).
\label{eq_op_FP-Sec1-1}
\end{align}
Because of the specific symmetry  exhibited by the Fokker--Planck operator (\ref{eq_op_FP-Sec1-1}) \cite{Fogedby12,Fogedby14}, the steady state PDF $P(Y)=\lim_{t\to \infty} P(Y,t)$ exhibits a long time fluctuation relation 
\begin{align}
P(Y)=P(-Y)\E^{\beta f Y}.
\label{eq_FT_Sec1-4}
\end{align}
As a consequence, the scaling cumulant generating function defined as~\cite{Touchette09} 
\begin{align}
\mu_0(\lambda)\equiv \lim_{t\to\infty} \frac{ \ln\langle\E^{\lambda Y}\rangle} t,
\label{eq_SCGF-Sec1-4}
\end{align}
that corresponds to the largest eigenvalues of the  operator (\ref{eq_op_FP-Sec1-1}),
exhibits the following symmetry  \cite{Fogedby12,Fogedby14}
\begin{equation}
\mu_0(\lambda)=\mu_0(-\lambda-f/T).
\label{sym:mu}
\end{equation}

We next introduce the generating function
\begin{align}
\Psi(y,\lambda,t)=\int_{-\infty}^{+\infty}\D Y\,\exp(\lambda Y)P(y,Y,t),
\label{eq_gen_func-Sec1-1}
\end{align}
whose time evolution $\partial_t \Psi= \hat {\bf L}_{\lambda} \Psi$ is governed by the differential operator 
\begin{align}
\hat {\bf L}_{\lambda}=-\partial_y\left(f-U_0'(y)-T\partial_y\right)+(f-U_0'(y))\lambda+T\lambda^2-2T\lambda\partial_y,
\label{eq_op_gen_func-Sec1-1}
\end{align}
which is a simplified version of the operator (\ref{eq_op_FP-Sec1-1}) as discussed in Ref.~\cite{Lebowitz1999}.
Considering the separation ansatz for $\Psi(y,\lambda,t)$ (Sec. 5.4 in~\cite{Risken}),
\begin{align}
\Psi(y,\lambda,t)=\varphi(y,\lambda)\E^{\mu(\lambda)t},
\label{eq_ansatz-Sec1-1}
\end{align}
one obtains 
\begin{align}
\hat {\bf L}_{\lambda}\varphi_n(y,\lambda)=\mu_n(\lambda)\varphi_n(y,\lambda),
\label{eq_eigenvalues_L-Sec1-1}
\end{align}
where  $\varphi_n(y,\lambda)$ and $\mu_n(\lambda)$ are the eigenfunctions and eigenvalues of the Fokker--Planck operator $\hat {\bf L}_{\lambda}$, respectively. To solve Eq.~\eqref{eq_eigenvalues_L-Sec1-1} we express the operator $\hat {\bf L}_{\lambda}$ Eq.~\eqref{eq_op_gen_func-Sec1-1} in matrix form, hence we need a complete and orthonormal basis, $\langle j| l\rangle=\delta_{jl}$. Because of the periodic nature of the system, a suitable choice for the basis is~\cite{Mehl08}
\begin{align}
\langle j|y\rangle=\frac{e^{-\ii  jy}}{\sqrt{L}},~\langle y|j\rangle=\frac{e^{\ii jy}}{\sqrt{L}}. 
\end{align}
 Expanding the eigenfunctions $\varphi_n(y,\lambda)$ into the chosen basis
\begin{align}
\varphi_n(y,\lambda)=\langle y|\varphi_n(\lambda)\rangle=\langle y|\sum_l c^{(n)}_l(\lambda)|l\rangle=\sum_l c^{(n)}_l(\lambda)\langle y|l\rangle,
\label{eq_exp_basis-Sec1-1}
\end{align}
equation~\eqref{eq_eigenvalues_L-Sec1-1} for the eigenvalue $n$ reads
\begin{align}
\sum_j L_{lj}c_j^{(n)}(\lambda)=\mu_n(\lambda)\, c_l^{(n)}(\lambda),
\label{eq:c}
\end{align}
with $~L_{lj}\equiv\langle l|\hat {\bf L}_{\lambda}|j\rangle=\int_0^{L} \D y\,\langle l|y\rangle \hat {\bf L}_{\lambda}\langle y|j\rangle$. 

Considering a cosine potential $U_0(y)=-k \cos y$ the matrix turns out to be tridiagonal with elements
\begin{eqnarray}
L_{jj}&=&-T(j+\ii \lambda)^2-if(j+\ii \lambda),~\text{if {\it j=l\,};}\label{eq_el_matrix_nn-Sec1-4}\\
L_{j,j\pm1}&=&\mp \frac{k}{2}\left(j+ \ii \lambda\right),~\text{if {\it j-l$=\pm 1$};}\label{eq_el_matrix_n1-Sec1-4}\\
L_{j l}&=&0,~\text{if $j\neq l$ and $j-l\neq\pm 1$.}\label{eq_el_matrix_nl-Sec1-1}
\end{eqnarray} 

\subsection{Perturbative approach}
\label{Sec1-2}
In its matrix form, the operator $\hat {\bf L}_{\lambda}$ Eq.~\eqref{eq_op_gen_func-Sec1-1} is an infinite matrix, whose size can be truncated to some finite value in order to solve the linear system Eq.~\eqref{eq:c}. We write $\hat {\bf L}_{\lambda}$ as the sum of a diagonal matrix and another one including the upper and the lower diagonals, $\hat {\bf L}_{\lambda}=\hat {\bf L}_{\lambda}^{(0)}+k\,\hat {\bf L}_{\lambda}^{(1)}$. The latter turns out to be proportional to the potential strength $k$, and therefore a perturbation theory can be used to obtain the eigenvalues as series expansions in terms of $k$: $\mu_n(\lambda)=\mu_n(\lambda)+k \mu_n^{(1)}(\lambda)+k^2 \mu_n^{(2)}(\lambda)+k^3 \mu_n^{(3)}(\lambda)+k^4 \mu_n^{(4)}(\lambda)+ O(k^5)$. 

However, the perturbation theory employed in quantum mechanics to solve, e.g.,  the Schr\"odinger equation, cannot be used here since neither the operator $\hat {\bf L}_{\lambda}$ of the Fokker--Planck equation nor the unperturbed operator $\hat {\bf L}_{\lambda}^{(0)}$ are Hermitian. Therefore, there is no set of functions to form a complete orthonormal basis for $\hat {\bf L}_{\lambda}^{(0)}$, and so the  corrections to the eigenvectors cannot be calculated.

One possible way to proceed in order to avoid this limitation is to recast the equation for the characteristic polynomial into the following form 
\begin{align}
\det\left[\hat {\bf L}_{\lambda}-\mu(\lambda)\mathbb{1}\right]=0\Rightarrow \det\left[{\bf M}\right]\,\det\left[\left(\mathbb{1}+k\,{\bf M}^{-1}\,\hat {\bf L}_{\lambda}^{(1)}\right)\right]=0,
\label{eq_recast-Sec1-2}
\end{align}
where $\hat {\bf L}_{\lambda}^{(0)}$ is a diagonal matrix with entries $L_{jj}$, ${\bf M}=\hat {\bf L}_{\lambda}^{(0)}-\mu(\lambda)\mathbb{1}$, and where we have employed the property that the determinant for the product of matrices is the product of their determinants. The matrix ${\bf M}$ is diagonal and thus its determinant reduces to the product of its entries: it depends on $\mu(\lambda)$ and can thus be easily expanded in powers of $k$. The expansion of the second determinant in Eq.~\eqref{eq_recast-Sec1-2} requires additional analysis that is included in~\appref{A1}.

\subsection{Largest eigenvalue}
\label{Sec1-3}
According to Eq.~\eqref{eq_ansatz-Sec1-1},  the long--time limit  behavior of the generating function $\Psi(y,\lambda,t)$ is dominated  
 by the largest eigenvalue of the operator $\hat {\bf L}_{\lambda}$,  which corresponds to the cumulant generating function $\mu_0(\lambda)$ introduced in Eq.~\eqref{eq_SCGF-Sec1-4}, and so we can write the generating function for $Y$ as 
\begin{align}
\psi(\lambda,t)=\int_0^L \D y\,  \Psi(y,\lambda,t) \sim\exp[\mu_0(\lambda)\,t].
\label{eq_func_gen_lim_t}
\end{align}
Thus, starting from the zeroth order, we apply the perturbative approach described above for the largest eigenvalue (labeled by $n=0$) to obtain the expansion for the largest eigenvalue up to fourth order,
\begin{equation}
\begin{split}
\mu_0(\lambda)\approx\quad\lambda\left(f+T\lambda\right)& \Biggl[1-\frac{k^2}{2}\frac{1}{T^2+(f+2T\lambda)^2}\\
& +\frac{k^4}{8}\frac{-f^4-3f^3 T \lambda+f^2 T^2(4+\lambda^2)+f T^3\lambda(9+8\lambda^2)+T^4(5+9\lambda^2+4\lambda^4)}{(T^2+(f+2T\lambda)^2)^3(4 T^2+(f+2T\lambda)^2)}\Biggr]+\mathcal{O}[k^6]\,.
\end{split}
\label{eq_mu0-Sec1-3}
\end{equation}
We recall that this last result is specific for the cosine potential $U_0(y)=-k \cos y$.

It is worth noting that by substituting the leading contribution of the generating function Eq.~\eqref{eq_func_gen_lim_t} into Eq.~\eqref{eq_gen_func-Sec1-1}, deriving with respect to $\lambda$, and evaluating the result at $\lambda=0$, we obtain the following identity
\begin{align}
\partial_{\lambda}\mu_0(\lambda)|_{\lambda=0}=\frac{\langle Y\rangle}{t}= \bar v_y.
\label{eq_vss_mu0-Sec1-3}
\end{align}
The last equation, together with Eq.~\eqref{eq_mu0-Sec1-3}, provides thus an expansion of the steady--state velocity $\bar v_y$ in powers of $k$. An identical result for $\bar v_y$ is obtained by considering  Eq.~\eqref{eq_ss_vel-Sec1} and expanding the normalization constant, as given by  Eqs.~\eqref{eq_pdf_ss-Sec1} and~\eqref{eq_normalization-Sec1}, in powers of $k$~\cite{Imparato15}.

\begin{figure}[h]
\center
\psfrag{D}[ct][ct][1.]{$D/D_0$}
\psfrag{f}[ct][ct][1.]{$f$}
\includegraphics[width=8cm]{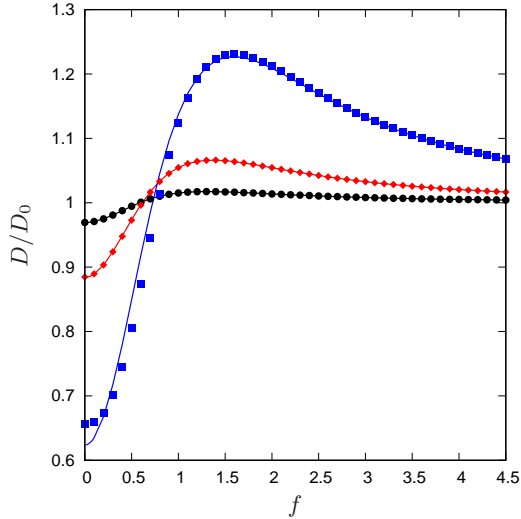}
\caption{Diffusion coefficient $D$ versus the tilt force $f$ for the single particle in the potential $U_0(y)=-k \cos y$. Full symbols:  $D$ as given by Eqs.~\eqref{eq_mu0-Sec1-3} and \eqref{eq_D_SCGF-Sec1-3}. Lines: the analytic prediction as given in~\cite{ReimannVandenBroeck01}. The system parameters are   $T=1,$ and $k=0.25$ (black circles), $k=0.5$ (red diamonds), $k=1$ (blue squares). $D_0=T$.}
\label{figure_D_Reimann_mu0}
\end{figure}

Further, the diffusion coefficient can be computed from $\mu_0(\lambda)$,
\begin{align}
D=\lim_{t\to \infty} \frac{\average{Y^2}-\average{Y}^2}{2 t}=\frac{1}{2}\partial_{\lambda}^2\mu_0(\lambda)|_{\lambda=0}.
\label{eq_D_SCGF-Sec1-3}
\end{align}
For small $k$,~\figref{figure_D_Reimann_mu0}, there is a good agreement between  the diffusion coefficient as given by  Eqs.~\eqref{eq_mu0-Sec1-3} and \eqref{eq_D_SCGF-Sec1-3}, and the analytic result calculated for the overdamped Brownian motion in a tilted periodic potential as obtained in~\cite{ReimannVandenBroeck01}.

\section{Two coupled oscillators}
\label{Sec2}
A minimal model for a machine with an input and an output energy current consists of two overdamped Brownian particles with coordinates $x_1$ and $x_2$  coupled through a periodic potential $U_0(x_1-x_2)$ of strength $k$ \cite{Golubeva2012,IzumidaKori16}. Each particle is subject to a constant tilting force of opposite sign, so that one injects energy into the system $f_1>0$, whereas the other extracts energy, $f_2<0$. The two particles will be termed the producer and the user, respectively. In the limit of weak coupling ($k$ small) the two particles tend to  move independently, each one at its own ``natural frequency" $f_i$, while strengthening $k$ will increasingly synchronize their motion. The dynamic equations for the two coupled oscillators read,
\begin{align}
\dot x_1=f_1-\partial_{x_1}U_0(x_1-x_2)+\zeta_1(t),\label{eq_langevin_x1-Sec2}\\
\dot x_2=f_2-\partial_{x_2}U_0(x_2-x_1)+\zeta_2(t)\label{eq_langevin_x2-Sec2}.
\end{align}
We assume uncorrelated Gaussian white noises, $\langle\zeta_i(t)\,\zeta_j(t')\rangle=2 T\delta_{ij}\delta(t-t')$, $i,j=1,2$. The efficiency Eq.~\eqref{eq_eff-Intro} for a single trajectory of this isothermal engine is the rate between the work Eq.~\eqref{eq_work-Sec1-1} extracted by the user along an individual trajectory and the work injected by the producer along the same trajectory 
\begin{align}
\eta=-\frac{w_2}{w_1}=-\frac{f_2\,X_2}{f_1\,X_1}.
\label{eq_eff}
\end{align}
We employ the same notation as in~\secref{Sec1-1} to distinguish the unbounded coordinates, $X_i$, from the bounded periodic ones $x_i$. The PDF of the efficiency is then~\cite{VerleyWillaert14},
\begin{align}
P(\eta,t)=\int^{+\infty}_{-\infty} \D X_1\int^{+\infty}_{-\infty} \D X_2\, P(X_1,X_2,t)\, \delta\left(\eta+\frac{f_2 X_2}{f_1X_1}\right).
\label{eq_PDF_eff-Sec2}
\end{align}
We need thus to evaluate the joint PDF $P(X_1,X_2,t)$ for the unbounded coordinates, that amounts to find the steady state solution to  the FP equation corresponding to the Langevin equations~\eqref{eq_langevin_x1-Sec2}-\eqref{eq_langevin_x2-Sec2}. To simplify the question, we introduce a coordinate transformation to recast the problem into the center of mass (CM) and the relative coordinate motion
\begin{align}
x\equiv\frac{x_1+x_2}{2},~y\equiv\frac{x_1-x_2}{2},
\label{eq_canvi_variables-Sec2}
\end{align}
so that the dynamic equations~\eqref{eq_langevin_x1-Sec2}-\eqref{eq_langevin_x2-Sec2} decouple into
\begin{eqnarray}
\dot x&=&f_x+\zeta_x(t),\label{eq_Langevin_x-Sec2}\\
\dot y&=&f_y-\partial_y U_0(2 y)/2+\zeta_y(t),\label{eq_Langevin_y-Sec2}
\end{eqnarray}
where $f_x\equiv(f_1+f_2)/2$, $f_y\equiv(f_1-f_2)/2$, and $\langle\zeta_{\alpha}(t)\,\zeta_{\beta}(t')\rangle=T\delta_{\alpha\beta}\delta(t-t')$.

The Langevin equation for the CM Eq.~\eqref{eq_Langevin_x-Sec2} describes unidimensional overdamped Brownian motion subject to a constant force $f_x$. Its PDF is hence a Gaussian centered at $f_x\,t$ (Eq. (5.20)~\cite{Risken})
\begin{align}
P(X,t)=\frac{1}{\sqrt{2\pi Tt}}\exp\left(-\frac{(X-f_x t)^2}{2Tt}\right),
\label{eq_px-Sec2}
\end{align}
where for consistency with our previous notation  we have introduced the  coordinate $X$ to remark that it is an unbounded degree of freedom. On the other hand, the Langevin equation for the relative coordinate Eq.~\eqref{eq_Langevin_y-Sec2} is analogous to Eq.~\eqref{eq_Langevin-Sec1} introduced in~\secref{Sec1} to discuss the  single oscillator. Thus we will exploit the results contained in that section to evaluate the PDF for the unbounded relative coordinate $P(Y,t)$. Unless otherwise indicated, we will assume the long-time limit, so that $y$ and $Y$ are uncorrelated, and we can make use of the separation ansatz Eq.~\eqref{eq_ansatz-Sec1-1}. In the following we will discuss different scenarios and approximations to obtain the PDF $P(Y,t)$.

However, before proceeding to analyze the PDF $P(Y,t)$, we consider the PDF of the efficiency Eq.~\eqref{eq_PDF_eff-Sec2}, and discuss a few simplifications. According to the transformation Eq.~\eqref{eq_canvi_variables-Sec2}, we have that $P(X_1,X_2,t)\propto P(X,t)P(Y,t)$ up to a constant given by the Jacobian of the coordinate transformation, hence the PDF of the efficiency Eq.~\eqref{eq_PDF_eff-Sec2} reads
\begin{align}
P(\eta,t)=\frac{1}{2}\int^{+\infty}_{-\infty} \D X\int^{+\infty}_{-\infty} \D Y\, P(X,t)\,P(Y,t)\, \delta\left(\eta+\frac{f_2 (X-Y)}{f_1(X+Y)}\right).
\label{eq_PDF_eff_XY-Sec2}
\end{align}

Let us introduce the rescaled trajectory dependent efficiency $ \hat \eta=-f_1\eta/f_2$, and the new variable
\begin{equation}
\xi=\frac X Y=\frac{1+\hat \eta}{1-\hat \eta}.
\label{eq_xi-Sec2}
\end{equation} 
Its PDF $\Phi(\xi,t)$ is such that the following general relation between the two probability distributions holds,
\begin{equation}
P(\eta,t)=\left| \frac{d \xi}{d \eta}\right| P(\xi,t)=\left| \frac{f_1}{f_2}\right| \frac 2 {(1-\hat \eta)^2}  \Phi(\xi,t).
\label{eq_P_eff-Sec2}
\end{equation}
Eq.~\eqref{eq_P_eff-Sec2} is a central result as it indicates that any large deviation contribution to $P(\eta,t)$ will arise from $\Phi(\xi,t)$.
The PDF of the new variable $\xi$ reads thus
\begin{equation}
 \Phi(\xi,t)=\int \D X \D Y \delta\p{\xi-\frac{X}{Y}} P(X,t) P(Y,t)=\int \D Y  \frac{|Y|}{\sqrt{2 \pi T t} } \E^{-
\frac{t (J_Y\xi -f_x )^2}{2 T}} P(Y,t),
\label{p:xi-Sec2}
\end{equation}
where the rhs of the equation is obtained by substituting the PDF for $X$ Eq.~\eqref{eq_px-Sec2}, and by introducing the rescaled variable $J_Y\equiv Y/t$, corresponding to the current associated to $Y$.

\subsection{Efficiency distribution with Gaussian approximation for the variable $Y$}
\label{Sec2-3}
The first case we consider is when $P(Y,t)$ is a Gaussian distribution. The assumption that the fluxes are normally distributed is commonly made when studying the thermodynamic properties of microscopic devices in the linear regime \cite{Verley2014, Polettini15, ProesmansDreher16, Polettini17}.
As argued below, the Gaussian approximation is accurate in the limit of 
 small $f_y$. Given the definition of $f_y=(f_1-f_2)/2$, this  translates into the requirement that the system is close to equilibrium, since the two forces $f_1$ and $f_2$ need to have opposite sign in order to constitute a duo of input/output power sources.

\begin{figure}[h]
\center
\psfrag{D}[ct][ct][1.]{$D/D_0$}
\psfrag{f}[ct][ct][1.]{$f_y$}
\includegraphics[width=16cm]{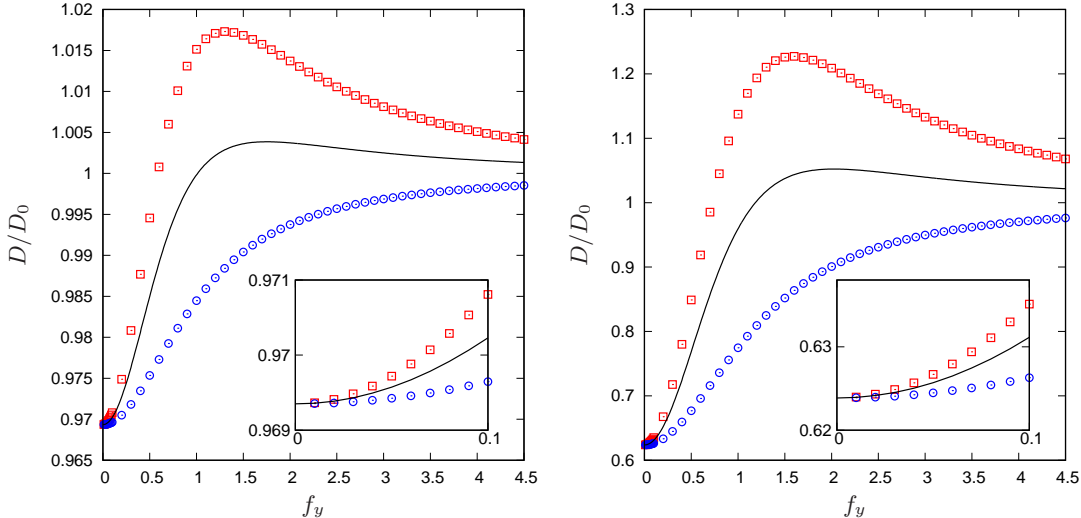}
\caption{Diffusion coefficient $D$ of the relative coordinate $Y$ versus the tilt force $f_y$ for the interaction potential  $U_0(y)=-2k \cos y$. Left panel: weak coupling  with $D_0=T=1$ and $k=0.25$. Right panel: larger coupling with $D_0=T=1$ and $k=1$. Blue circles: $D$ as obtained by Eq.~\eqref{eq_D-Sec2-3} (Gaussian approximation). Red squares: exact result as obtained in~\cite{ReimannVandenBroeck01}. Full black line: analytic approximation $D=T\D\langle\dot y\rangle/\D f_y$ as obtained in~\cite{CostatiniMarchesoni99}.}
\label{figure_D}
\end{figure}

Taking $P(Y,t)$ to be Gaussian corresponds to truncate the 
cumulant generating function $\mu_0(\lambda)$ to second order in $\lambda$. While the first order coefficient is fixed by the average value of the velocity, Eq.~\eqref{eq_vss_mu0-Sec1-3}, the second order coefficient is dictated by the symmetry imposed by the fluctuation relation $\mu_0(\lambda)=\mu_0(-\lambda-f_y/T_y)$ Eq.~\eqref{sym:mu}, with $T_y=T/2$ because of the coordinate change in Eq.~\eqref{eq_canvi_variables-Sec2}.
The cumulant generating function of $Y$ for the Gaussian approximation thus reads 
\begin{equation}
\mu_0(\lambda)=\bar v_{y}(k,f_y)\lambda(1+ \lambda T_y/f_y ), 
\label{m0gaus1}
\end{equation}
where $\bar v_{y}(k,f_y)=\langle\dot Y\rangle$ is given by Eq.~\eqref{eq_ss_vel-Sec1}.
The expression~\eqref{m0gaus1} for $\mu_0(\lambda)$ sets a constraint on the diffusion coefficient in the Gaussian distribution $P(Y,t)$ that reads 
\begin{equation}
P(Y,t)=\exp\pq{-\frac{f_y (Y -\bar v_{y} t )^2}{2t  T \bar v_{y}}}\frac{1}{\sqrt{2 \pi t  T \bar v_{y} /f_y}},
\label{pygaus1-Sec2-3}
\end{equation}
where we have dropped the dependency of $\bar v_{y}$ on $k$ and $f_y$ to simplify the notation.
The same result is obtained if one imposes directly the fluctuation relation Eq.~\eqref{eq_FT_Sec1-4} on a Gaussian distribution with average $\bar v_y t$.
In order to check the accuracy of the Gaussian approximation Eq.~\eqref{pygaus1-Sec2-3}, we compare the diffusion coefficient as given by our approximation for the PDF Eq.~\eqref{pygaus1-Sec2-3},
\begin{align}
D=\frac{\sigma_Y^2}{2t}=\frac{T \bar v_y(k,f_y)}{2 f_y},
\label{eq_D-Sec2-3}
\end{align}
with the exact result obtained in~\cite{ReimannVandenBroeck01}. In the limit of small force $f_y$ the diffusion coefficient $D$  Eq.~\eqref{eq_D-Sec2-3} obtained  with the Gaussian approximation 
agrees with the analytic expression as obtained in~\cite{ReimannVandenBroeck01}, see~\figref{figure_D}, and in particular its insets.  Yet the diffusion coefficient in Eq.~\eqref{eq_D-Sec2-3} does not exhibit the peak near the ``critical tilt". Besides, the degree of agreement is improved in the weak coupling regime, notice the different scales employed in the vertical axes of the inset plots in~\figref{figure_D}.

\begin{figure}[h]
\center
\psfrag{x}[ct][ct][1.]{$\eta$}
\psfrag{y}[cc][cc][1.]{$P(\eta,t)$}
\includegraphics[width=18cm]{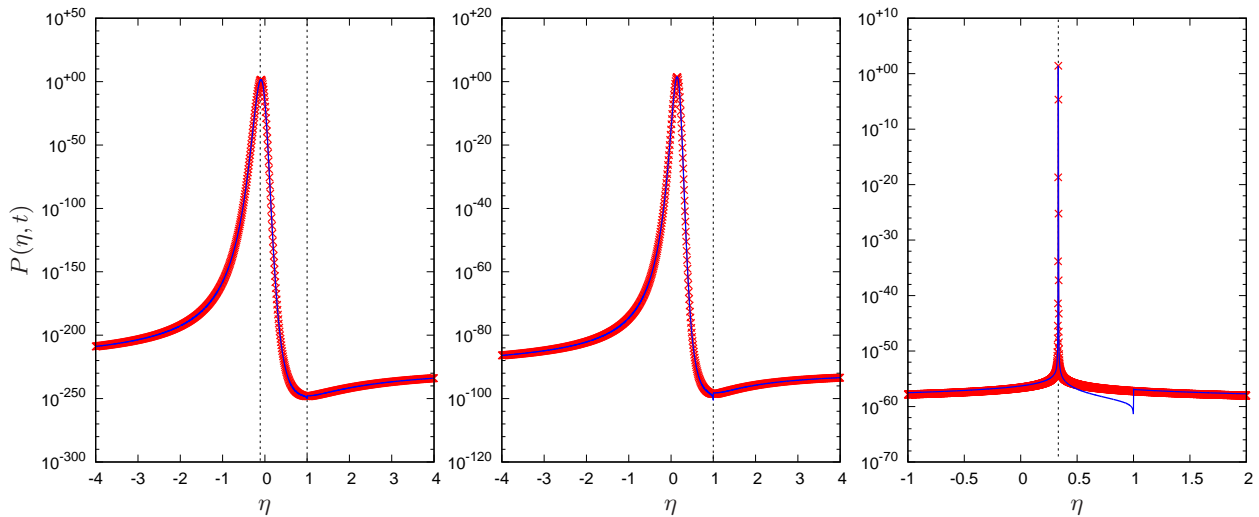}
\caption{PDF of the efficiency for weak coupling ($k=0.25$, left panel), moderate coupling ($k=1$, middle panel), and tight coupling ($k=4$, right panel). Red symbols: exact expression for $P(\eta,t)$ Eq.~\eqref{eq_P_eta-Sec2-3}. Blue line: long time approximation for $P(\eta,t)$ Eq.~\eqref{eq_P_eta_t-Sec2-3}. Vertical black lines: extremal points of $P(\eta,t)$ Eq.~\eqref{eq:extr}, Eq.~\eqref{limk0} (left panel), Eq.~\eqref{limkinf} (right panel). Parameters choice: $T=1,f_x=0.05,f_y=0.1,t=10^{5}$. The average velocity $\bar v_y$ appearing in Eqs.~\eqref{eq_P_eta-Sec2-3} and~\eqref{eq_P_eta_t-Sec2-3} is calculated through the exact expression \eqref{eq_ss_vel-Sec1}, for the interaction potential $U_0(y)=-2k\,\cos y$.}
\label{figure_P_eta}
\end{figure}

We now proceed to calculate the PDF $P(\eta,t)$. 
Plugging the expression for $P(Y,t)$ Eq.~\eqref{pygaus1-Sec2-3} into Eq.~\eqref{p:xi-Sec2}, integrating over $Y$, and inverting the change of variables Eq.~\eqref{eq_xi-Sec2}, we obtain
\begin{align}
P(\eta,t)=\E^{-\frac{t(f_x^2+f_y \bar v_y)}{2T}}\frac{4 f_y T\, h(\eta)^2}{(f_x+f_y)\pi\sqrt{f_y \bar v_y}(\eta-1)^2|f_x-f_y|}\left[1+\sqrt{\pi t}\,h(\eta)\,\E^{t\, h(\eta)^2}\erf{\left(\sqrt{t}\, h(\eta)\right)}\right],
\label{eq_P_eta-Sec2-3}
\end{align}
where $h(\eta)=(f_x^2-f_y^2)(\eta-1)\sqrt{\bar v_y}\left((f_y(\eta-1)+f_x(\eta+1))\sqrt{2T\left(f_y+\frac{\bar v_y(f_x(\eta-1)+f_y(\eta+1))^2}{(f_y(\eta-1)+f_x(\eta+1))^2}\right)}\right)^{-1}$, see ~\appref{A2} for the details. Finally, in the long time limit the leading terms of the efficiency's PDF are
\begin{align}
P(\eta,t)=\E^{-\frac{t(f_x^2+f_y \bar v_y)}{2T}}\frac{4 f_y T\, h(\eta)^2}{(f_x+f_y)\pi\sqrt{f_y \bar v_y}(\eta-1)^2|f_x-f_y|}\left[1-\frac{h(\eta)}{|h(\eta)|}+\sqrt{\pi t}\,|h(\eta)|\,\E^{t\,h(\eta)^2}\right],
\label{eq_P_eta_t-Sec2-3}
\end{align}
and they exhibit a good agreement with the exact expression Eq.~\eqref{eq_P_eta-Sec2-3}, see~\figref{figure_P_eta}.
The long time approximation Eq.~\eqref{eq_P_eta_t-Sec2-3} exhibits a discontinuity at $\eta=1$ that fades when $\sqrt{t\bar v_y}$ is large. As such it can be observed in the tight coupling regime, i.e., when $\bar v_y$ is small, even at a large time, see~\figref{figure_P_eta} rightmost panel.

The efficiency distribution in~\figref{figure_P_eta} exhibits a maximum and a minimum, which correspond to the large deviation function's extremal points, as detailed below.
The super--Carnot local maximum found in~\cite{Polettini15} belongs to a subdominant decay mode, and thus it does not appear in the plot ranges of~\figref{figure_P_eta} as it is displaced towards infinity in the long time limit.
From Eq.~\eqref{eq_P_eta_t-Sec2-3} one obtains the large deviation function of the PDF of the efficiency
\begin{align}
J(\eta)\equiv\lim_{ t\to\infty} \frac{1}{t}\ln P_t(\eta),
\label{eq_ldvf_eta-Sec2-3}
\end{align}
that has two extremal points
\begin{equation}
\eta_+=\frac{f_2( c(k,f_1-f_2)+2 f_2)}{f_1(c(k,f_1-f_2)-2 f_1)};\qquad \eta_-=1;
\label{eq:extr}
\end{equation} 
which are a maximum and a minimum, respectively; where we have written the velocity $\bar v_y$ in the form of Eq.~\eqref{eq_ss_vel_c-Sec1} in~\secref{Sec1}. 
The minimum  $\eta_-=1$, corresponding to the reversible efficiency, is in accordance with the findings of \cite{Verley2014} on the least likely efficiency in stochastic machines, and is a direct consequence of the fluctuation relation Eq.~\eqref{eq_FT_Sec1-4}.
Furthermore, we find that the most likely value of the efficiency $\eta_+$ is always equal to the macroscopic efficiency $\bar \eta$ 
\begin{equation}
\eta_+=\bar \eta=-\frac {f_2\average{ X_2}} {f_1 \average{X_1}},
\label{etabar}
\end{equation} 
which, differently from the minimum, depends on the coupling strength $k$ and on the forces.
We will now consider the limiting cases of weak and tight coupling and of large forces. In this analysis we will avail ourselves of the results on the single oscillator velocity discussed in \secref{Sec1}.
 
For $k=0$, we have $\bar v_y(0,f_y)=f_y$ and thus
\begin{equation}
\lim_{k \to 0}\eta_+=-\p{\frac{f_2}{f_1}}^2.
\label{limk0}
\end{equation} 
The same result is obtained in the limit $f_y\to \infty$; indeed, in \secref{Sec1} we have found 
\begin{equation}
\lim_{f_y\to \infty} \bar v_y(k,f_y)=f_y,\; \forall k<\infty,
\label{eq_lim_large_f-Sec2-5}
\end{equation} 
and recalling that $f_y=(f_1-f_2)/2$, one finds
\begin{equation}
\lim_{f_y\to + \infty}\eta_+=\lim_{f_1\to+\infty} \eta_+=\lim_{f_2\to -\infty} \eta_+=-\p{\frac{f_2}{f_1}}^2.
\label{limy}
\end{equation} 
Thus we find that a large applied force renormalizes the interaction potential, leading to a non-interacting system with negative macroscopic efficiency.
It is worth noting that the minimum Eq.~\eqref{eq:extr} and the weak coupling maximum Eq.~\eqref{limk0} of the large deviation function $J(\eta)$ match the extremal points of the efficiency distribution as shown in \figref{figure_P_eta} (left panel) for a particular choice of the system parameters.

In the limit of tight coupling  $k\to \infty$ the variable $Y$ becomes confined, and  one has
\begin{equation}
\lim_{k\to \infty} \bar v_y(k,f_y)=0,\; \forall f_y<\infty,
\label{limk}
\end{equation} 
thus
\begin{equation}
\lim_{k\to\infty} \eta_+=  -\frac{f_2 }{f_1}.
\label{limkinf}
\end{equation}
This tight coupling limit can be seen in~\figref{figure_P_eta} (right panel).
We now argue that the values given in Eq.~\eqref{limk0} and  Eq.~\eqref{limkinf} are respectively the lower and the upper bounds for the most likely and thus for the macroscopic efficiency Eq.~\eqref{etabar}.
Indeed, one finds that
\begin{equation}
\partial_k \eta_+= \frac{2f_2 f_x}{f_1 (f_x+ \bar v_y(k,f_y))^2} \partial_k \bar v_y(k,f_y)>0,
\end{equation} 
where we have used the fact that $\partial_k \bar v_y(k,f_y)<0$ for $f_y>0$, and 
we have assumed that $f_1>0$, $f_2<0$, with $|f_1|>|f_2|$.
Furthermore, $-(f_2/f_1)^2<-f_2/f_1$, and so $\eta_+$ is restricted in this interval of values.
Thus we conclude that i) the optimal maximal/macroscopic efficiency is always obtained in the limit of tight coupling, where the relative coordinate $Y$ and its fluctuations are suppressed \cite{Imparato15}, and ii) the only way that the maximal/macroscopic efficiency can reach the reversible value 1 is in the limit $f_2\to -f_1$ for which the total entropy production in the environment vanishes. 
The latter result is relevant in connection with the argument raised in \cite{Polettini17}, where it was argued that a machine at diverging power output can achieve the reversible efficiency limit.  On the one hand, our results show clearly that, for finite coupling strength $k$, taking diverging $f_i$ gives the lower bound Eq.~\eqref{limy}. On the other hand, by taking first the tight coupling limit,  $k\to \infty$, and then $f_1$ and $f_2$ large but with $f_i\ll k$, the machine can achieve a large power output ($f_2 \average{\dot x_2}$), but at the expenses of a large power input ($f_1 \average{\dot x_1}$), given that $\bar v_y({k\to\infty},f_y)\to 0$ (Eq.~\eqref{limk}) and thus $\langle\dot x_2\rangle\to\langle\dot x_1\rangle$. Therefore, the macroscopic efficiency 1 can only be achieved close to the stall condition  $f_2\to -f_1$, making the machine a dud.

The thermodynamic uncertainty relation~\cite{BaratoSeifert15,GingrichHorowitz16} sets an upper bound for the thermodynamic efficiency, which reads~\cite{PietzonkaSeifert18}, 
\begin{align}
\bar\eta\leq \frac{1}{1+2\average{\dot w_2} T/\Delta_2},
\label{eq_Seifert}
\end{align}
where $\average{\dot w_2}$ is the average output power and $\Delta_2$ its fluctuations,
\begin{align}
\langle \dot w_2\rangle=\frac{-f_2 \langle X_2\rangle}{t},~\Delta_2=\lim_{t\to\infty}\langle(\dot w_2(t)-\average{\dot w_2})^2\rangle t.
\end{align}
By virtue of the coordinate transformation Eq.~\eqref{eq_canvi_variables-Sec2} and given the distribution of $X$ Eq.~\eqref{eq_px-Sec2}, the mean power and its fluctuations read,
\begin{align}
\average{\dot w_2}=-f_2\left(\frac{f_1+f_2}{2}-\bar v_y\right),~\Delta_2=\lim_{t\to\infty}\frac{f_2^2}{t}\left(Tt+\Delta_Y\right).
\end{align}
The steady--state mean velocity $\bar v_y$ has already been discussed in~\secref{Sec1} (Eq.~\eqref{eq_ss_vel_c-Sec1}), whereas the fluctuations of $Y$ decrease from $\Delta_Y=tT$ in the weak coupling limit ($k\to 0$) to $\Delta_Y=0$ in the tight coupling limit ($k\to \infty$, when the two oscillators are fully coupled and so the relative coordinate vanishes). We therefore find that, for decreasing $k$,
\begin{align}
0\leq\Delta_Y\leq tT.
\end{align}
According to this argument, the upper bound for the macroscopic efficiency ($f_1\neq -f_2$), as given by Eq.~\eqref{eq_Seifert},  takes the values 
\begin{align}
\bar\eta\leq \infty,~\text{for $k\to 0$,}\\
\bar\eta\leq \frac{-f_2}{f_1},~\text{for $k\to\infty$.}
\end{align}
Comparing these results with the asymptotic behaviors of the most likely efficiency Eqs.~\eqref{limk0} and~\eqref{limkinf}, we notice that the upper bound Eq.~\eqref{eq_Seifert} turns out to overestimate by far the  macroscopic efficiency in the weak coupling, whereas for in the tight coupling we prove that the upper bound corresponds to the actual value for the macroscopic efficiency.

\subsection{Saddle-point approach}
\label{Sec2-4}
We now calculate the probability distribution $\Phi(\xi,t)$, Eq.\eqref{p:xi-Sec2}, and the most and the least probable value of the efficiency without making any assumption on the relative coordinate distribution.
Recalling that $\mu_0(\lambda)$ introduced in Eq.~\eqref{eq_SCGF-Sec1-4} is the cumulant generating function of $Y$, we have that in the long time limit  
\begin{equation}
P(Y,t)\sim\int \D \lambda  \E^{t[\mu_0(\lambda)-\lambda \my]} \propto \E^{t[\mu_0(\ls(\my))-\ls(\my) \my]},
\label{eq_PDF_Y_LDF}
\end{equation} 
where $\ls$ is implicitly defined by the saddle--point condition 
\begin{equation}
 \left.\partial_\lambda\mu_0(\lambda)\right|_\ls=\my.
\label{eq_dmu-Sec2-4}
\end{equation}
The integral in Eq.~\eqref{p:xi-Sec2} is dominated by the saddle point $\myss$ defined implicitly by 
\begin{equation}
 \ls(\myss)=-\frac{\xi}{T} (\myss \xi-f_x),
\label{eq_yss-Sec2-4}
\end{equation} 
where Eq.~\eqref{eq_dmu-Sec2-4} is exploited to simplify the last expression.
Thus one obtains
\begin{align}
\Phi(\xi,t)\sim\E^{t G(\xi)}=\E^{-t [(\myss\xi -f_x )^2/2 T-(\mu_0(\lss)-\lss \myss)]},
\label{eq_P_xi-Sec2-4}
\end{align} 
with $\lss= \ls(\myss)$. Let us now find the stationary points of $G(\xi(\eta))$. We first notice that
\begin{align}
\frac{\partial G}{\partial\eta}=\frac{\partial\xi}{\partial\eta}\frac{\partial G}{\partial\xi}=-\frac{f_1}{f_2}\frac{1}{2}(1+\xi)^2\frac{\partial G}{\partial \xi}.
\label{eq_deta-Sec2-4}
\end{align}
Exploiting equations~\eqref{eq_dmu-Sec2-4}-\eqref{eq_yss-Sec2-4}, a straightforward calculation leads to the expression for the stationary points
\begin{align}
\frac{\partial G}{\partial\xi}=\frac{\lambda^{**}\my^{**}}{\xi}=0.
\label{eq_st-Sec2-4}
\end{align} 
Thus, we are left with the two equations,
\begin{equation}
\lss=0 ;\;\myss=0.
\end{equation} 
The first equation together with Eq.~\eqref{eq_yss-Sec2-4} gives
\begin{equation}
0=\xi(\myss\xi-f_x).
\end{equation}
The solution $\xi=0$ must be discarded, because $\xi$ appears in the denominator of Eq.~\eqref{eq_st-Sec2-4}, so we have
\begin{equation}
\xi_+=f_x/\bar v_y\Rightarrow  \eta_+=\frac{f_2 (\bar v_y-f_x)}{f_1 (f_x+\bar v_y)},
\label{eq_eta_max-Sec2-4}
\end{equation} 
where we have used Eq.~\eqref{eq_dmu-Sec2-4} and the fact that when $\lss=0$,
\begin{equation}
\myss= \left.\partial_\lambda\mu_0(\lambda)\right|_{\lss=0}=\bar v_y.
\end{equation} 
The other solution of Eq.~\eqref{eq_st-Sec2-4}, $\myss=0$, implies 
\begin{equation}
\left.\partial_\lambda\mu_0(\lambda)\right|_{\lss}=0,
\end{equation}  
so $\lss$ in this case is the minimum of $\mu_0(\lambda)$, which is a convex function. Since the fluctuation relation for the FP equation with operator given by Eq.~\eqref{eq_op_gen_func-Sec1-1} implies $\mu_0(\lambda)=\mu_0(-\lambda-2 f_y/T)$~\cite{Fogedby12}, and thus $\mu_0(\lambda)$ is symmetric around $\lambda=-f_y/T$, we have that the minimum is exactly at this symmetry point. Hence, $\lss=-f_y/T$, and thus exploiting Eq.~\eqref{eq_dmu-Sec2-4} the least likely $\xi$ and $\eta$ are 
\begin{equation}
\xi_-=-f_y/f_x\Rightarrow \eta_-=1.
\label{eq_eta_min-Sec2-4}
\end{equation} 
Therefore, the solutions for $\eta_+,\, \eta_- $ are the same as those in Eq.~\eqref{eq:extr}, obtained with the Gaussian approximation for the current $J_Y$.

\subsection{ Linear regime and singular coupling}
\label{Sec2-6}
After the usual coordinate transformation into the CM and the relative coordinate Eq.~\eqref{eq_canvi_variables-Sec2}, the average velocities for the two oscillators read
\begin{align}
\bar v_1=\langle \dot x_1\rangle=\langle \dot X\rangle+\langle \dot Y\rangle=f_x+\bar v_{y}\left(f_y,k\right),\\
\bar v_2=\langle \dot x_2\rangle=\langle \dot X\rangle-\langle \dot Y\rangle=f_x-\bar v_{y}\left(f_y,k\right),
\end{align}
with $f_{x,y}=(f_1\pm f_2)/2$, and where we notice that in order for the machine to extract work from the input source of power, the two forces must be of opposite sign as discussed above.
We consider the linear regime between fluxes (particle velocities) and thermodynamic forces  $f_1$ and $f_2$,
\begin{align}
\bar v_i= L_{ij}f_j\,,
\label{eq_response_matrix}
\end{align}
where the linear response matrix is
\[
{\bf L}=\frac 1 2 \begin{bmatrix}
 1  +b(k) & 1  -b(k) \\
 1  -b(k)&  1  +b(k) \\
\end{bmatrix},
\]
and
\begin{align}
b(k)=\frac{\partial \bar v_{y}}{\partial f_y}\Big|_{f_y= 0}=I_0^{-2}\left(k\right)\,,
\label{eq_partial_vss}
\end{align}
is the partial derivative of the steady state velocity Eq.~\eqref{eq_ss_vel-Sec1} calculated  at $f_y=0$; 
$I_0\left(k\right)=\frac{1}{2\pi}\int_0^{2\pi}dy\,\exp\left(2\beta\,k\,\cos y\right)$ is the zeroth order modified Bessel function of the first kind.
In deriving the expression for ${\bf L}$ we have assumed that since  $f_1$ and $f_2$ are both small, their difference $2f_y$ is small as well, and the interacting potential $U_0(y)=-2k \cos y$. 

The function $b(k)$ is monotonically decreasing from $b(0)=1$ to $b(k\to\infty)\to 0$. Thus, in the linear regime
 the macroscopic efficiency $\bar \eta$ achieves the reversible limit 1 only in the limit of tight coupling $k\to \infty$.
This result is in agreement with the analysis we discussed in \secref{Sec2-3} for the general case of arbitrary forces $f_1$ and $f_2$ in the tight coupling limit, as summarized by Eq.~\eqref{limkinf}.
In the same section we derived the range of values for the  macroscopic efficiency by using a general argument. Obviously, the values of the macroscopic efficiency are limited in that range in the linear regime too.
It is however interesting to investigate whether in the linear regime one can attain the condition called {\it  singular coupling} in \cite{Polettini15}, where the reversible efficiency can be achieved when the linear response matrix tends to the inverse of a degenerate matrix.
The entries of the inverse matrix of $\bf L$ are
\begin{align}
 L_{ii}^{-1}=\frac 1 {b(k)}  L_{ii};\label{eq_inv_response_matrix1-Sec2-6}\\
 L_{ij}^{-1}=-\frac 1 {b(k)}  L_{ij},\,i\neq j .
\label{eq_inv_response_matrix2-Sec2-6}
\end{align} 
Such a matrix becomes degenerate in the limit $b(k)\to \infty$, which is not a physically meaningful limit: the response of a current (in our case the derivative of $\bar v_y$, i.e. the particle current) cannot be infinite for any finite value of the corresponding thermodynamic force ( in our case $f_y$).
 Therefore, when one considers a physical model for an engine, with realistic physical interaction between the thermodynamic forces, and thus between the corresponding energy currents, the necessary (but not sufficient) condition for the engine to operate at a macroscopic efficiency near the reversible (Carnot efficiency) is that the coupling between the input and output currents is tight.

\subsection{Fluctuation Theorem for the efficiency PDF}
\label{Sec2-7}
\begin{figure}[h]
\center
\psfrag{x}[ct][ct][1.]{$\eta$}
\psfrag{y}[cc][cc][1.]{$P(\eta,t)$}
\psfrag{x1}[ct][ct][1.]{$\eta$}
\psfrag{y1}[cc][cc][1.]{$\eta'$}
\includegraphics[width=16cm]{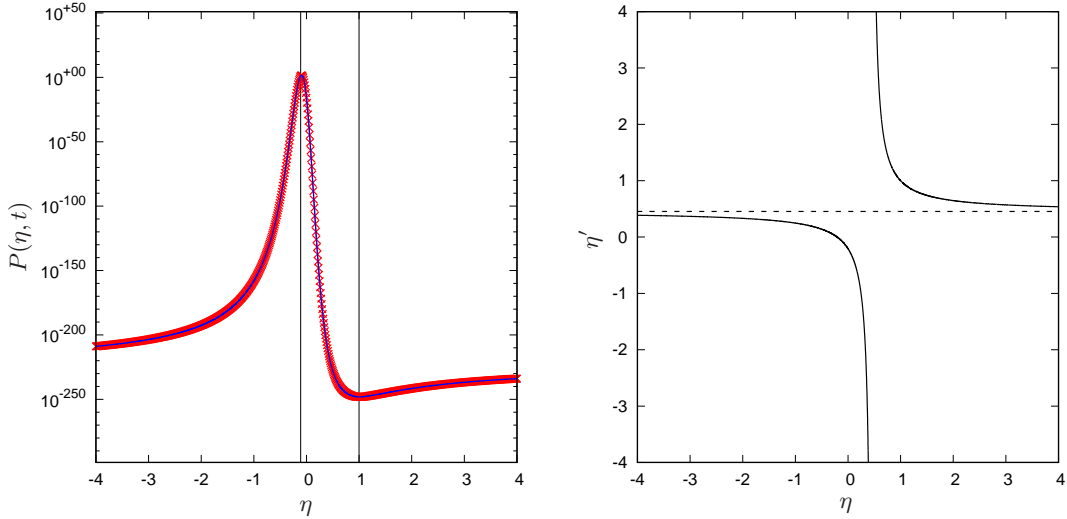}
\caption{Left panel: fluctuation symmetry for the efficiency PDF. Red symbols: exact expression for $P(\eta,t)$ Eq.~\eqref{eq_P_eta-Sec2-3}.  The average velocity $\bar v_y$ appearing in Eqs.~\eqref{eq_P_eta-Sec2-3}  is calculated through the exact expression \eqref{eq_ss_vel-Sec1}, for the interaction potential $U_0(y)=-2k\,\cos y$. Blue full line: transformation of $P(\eta,t)$ according to the fluctuation relation Eq.~\eqref{FT_eta}. Vertical black lines: extremal points of $P(\eta,t)$ Eq.~\eqref{eq:extr} and Eq.~\eqref{limk0}. Right panel, full line: $\eta'$ as a function of $\eta$ as given by Eq.~\eqref{eq_eta_etap}; the dashed line is the horizontal asymptote predicted by Eq.~\eqref{eq_eta_etap}. Parameter choice for both panels: $T=1,f_x=0.05,f_y=0.1,k=0.25,t=10^{5}$.}
\label{figure_P_eta_FT}
\end{figure}

The PDF of the position of  the single particle described by Eq.~\eqref{eq_Langevin-Sec1} exhibits the long time fluctuation relation  as given by 
Eq.~\eqref{eq_FT_Sec1-4}. We now explore whether the PDF of the efficiency $P(\eta,t)$ exhibits any fluctuation symmetry: according to Eq.~\eqref{eq_P_eff-Sec2} any symmetry in  $P(\eta,t)$  must correspond to a symmetry in $\Phi(\xi,t)$ as given by Eq.~\eqref{p:xi-Sec2}. Thus we would like to find a  transformation $g(\xi)$, such that
\begin{equation}
\Phi(\xi,t)\propto \Phi(g(\xi),t)=\int \D Y  |Y|/(\sqrt{2 \pi T t}) \exp\left[{-\frac{t (\my g(\xi) -f_x )^2}{2 T}}\right] P(Y,t),
\label{eq_sym_sec2-7}
\end{equation}
 where we have set again $\my\equiv Y/t$. Assuming a Gaussian distribution for $P(Y,t)$ Eq.~\eqref{pygaus1-Sec2-3}, changing variable $\my=\alpha\my'$, and setting
\begin{eqnarray}
g(\xi)&=&\frac{2 f_y f_x-\xi (f_y   \bar v_y-  f_x^2)}{f_y \bar v_y-f_x^2+2 \xi  f_x \bar v_y},\label{eq:g}\\
\alpha(\xi)&=&\left|\frac{-f_y \bar v_y+f_x^2-2 \xi  f_x \bar v_y}{f_y \bar v_y+f_x^2}\right|\,,
\label{eq:a}
\end{eqnarray}
we find that the fluctuation relation for the stochastic variable $\xi$ reads
\begin{equation}
\Phi(\xi,t)= \Phi(g(\xi),t)\E^{-R(\xi)}\,,
\label{eq_FT_PDF_eff}
\end{equation}
with $R(\xi)=\ln \alpha^{2}(\xi)$.
This symmetry turns out be analogous to the fluctuation relations of the work or the heat PDFs \cite{Seifert2012,Imparato06,Ciliberto2013,Ciliberto2013a,Berut16,Berut16a}.
One finds that for any value of $f_x,\, f_y$ and $\bar v_y$, $\partial_\xi g(\xi)<0, \, \forall \xi\neq f_x^2/(2f_y \bar v_y)-1/2$ where the function $g(\xi)$ has a vertical asymptote. Thus the function $g(\xi)$ is biunivocal: for any value of $\xi$ there is one and only one corresponding value $g(\xi)$.
Interestingly, if we take the two stationary points Eqs.~\eqref{eq_eta_max-Sec2-4} and~\eqref{eq_eta_min-Sec2-4} we find $g(\xi_{\pm})=\xi_{\pm}$ and $\alpha(\xi_{\pm})=1$,
i.e., the maximum and the minimum of $\Phi(\xi,t)$ are mapped into themselves.
Recalling the definition of $\xi=X/Y$, the quantity $R(\xi)$ can be seen as a measure of the deviation of a given trajectory from the typical trajectories  leading to the extremal values of the efficiency $\xi_{\pm}$ (or $\eta_\pm$).  While $R(\xi)$ vanishes at such points, it diverges in the limit $\xi\to \pm \infty$.
By taking into account the relation between $\eta$ and $\xi$, Eq.~\eqref{eq_xi-Sec2}, and the relation between their PDFs, Eq.~\eqref{eq_P_eff-Sec2}, we obtain the somewhat convoluted fluctuation relation for $P(\eta,t)$,
\begin{eqnarray}
P(\eta,t)&=& \left| \frac{f_1}{f_2}\right| \frac 2 {(1-\hat \eta)^2}  \Phi(\xi(\eta),t)=\left| \frac{f_1}{f_2}\right| \frac 2 {(1-\hat \eta)^2} \E^{-R(\xi(\eta))} \Phi(g(\xi(\eta),t)),
\label{FT_Peta}
\end{eqnarray} 
which can be recast in the simpler form,
\begin{equation}
P(\eta,t)=\frac{(1-\hat \eta')^2}{(1-\hat \eta)^2}  \E^{-R(\xi(\eta))} P(\eta',t),
\label{FT_eta}
\end{equation} 
with $\eta'$ implicitly defined by the equation 
\begin{equation}
\xi(\eta')=g(\xi(\eta)).
\label{eq_eta_etap}
\end{equation} 
This fluctuation symmetry for $P(\eta,t)$ is depicted in the left panel of \figref{figure_P_eta_FT}.
The relation between $\eta$ and $\eta'$ is depicted in  \figref{figure_P_eta_FT} (right panel).


\begin{figure}[h]
\center
\psfrag{x}[ct][ct][1.]{$\xi$}
\psfrag{y}[cc][cc][1.]{$\Phi(\xi,t)$}
\includegraphics[width=18cm]{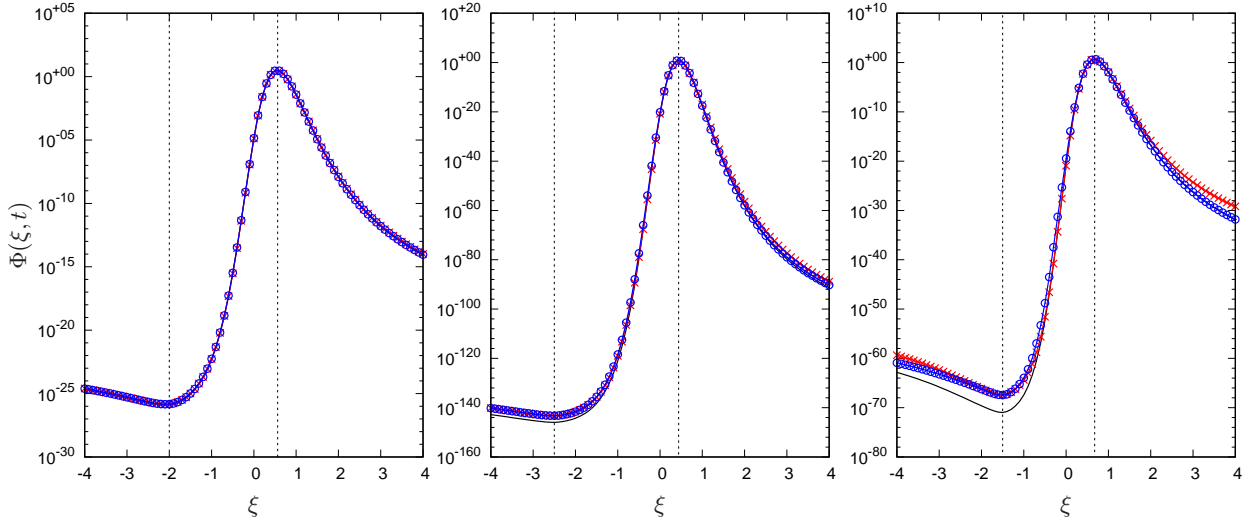}
\caption{Fluctuation symmetry for $\Phi(\xi,t)$  Eq.~\eqref{eq_FT_PDF_eff}. Black line: exact expression for $\Phi(\xi,t)$ (see \appref{A2}) assuming a Gaussian distribution for $Y$ Eq.~\eqref{pygaus1-Sec2-3}. Red crosses: $\Phi(\xi,t)$ as obtained by the numeric integration of Eq.~\eqref{p:xi-Sec2} with $P(Y,t)$ computed from the generating function Eq.~\eqref{eq_func_gen_lim_t}, and with the largest eigenvalue $\mu_0(\lambda)$ given by Eq.~\eqref{eq_mu0-Sec1-3}. The  mean velocity $\bar v_y$ is given by   Eq.~\eqref{eq_vss_mu0-Sec1-3}. Blue circles: transformed $\Phi(\xi,t)$, right hand side of Eq.~\eqref{eq_FT_PDF_eff}, as obtained with the Gaussian assumption for $P(Y,t)$. Dashed black lines: extremal points of $\Phi(\xi,t)$ Eqs.~\eqref{eq_eta_max-Sec2-4} and \eqref{eq_eta_min-Sec2-4}.
Parameter choice (the numbers  in parentheses refer to the values for each panel from left to right): $T=1,f_x=(0.05,0.1,1.0),f_y=(0.1,0.25,1.5),k=0.25,t=10^{5}$ (left, middle), $100$ (right panel).}
\label{figure_Phi_xi_FT}
\end{figure}

The fluctuation relations for the variables $\xi$ and $\eta$ discussed above were obtained under the assumption that the PDF of the relative variable $Y$ is the Gaussian function in Eq.~\eqref{pygaus1-Sec2-3}. As we argued above,~\secref{Sec2-3}, this approximation holds in the limit of small force $f_y$,~\figref{figure_D}.
However one might wonder whether the fluctuation relations Eqs.~\eqref{eq_FT_PDF_eff} and \eqref{FT_eta} still hold when one drops the assumption that $Y$ is Gaussian distributed.
To check this hypothesis we can exploit our result for the cumulant generating function up to the forth order in $k$ Eq.~\eqref{eq_mu0-Sec1-3}, obtained for the cosine potential, so as $P(Y,t)$ can be obtained through the saddle--point approximation Eq.~\eqref{eq_PDF_Y_LDF}.
The PDF $\Phi(\xi,t)$ and $\Phi(g(\xi),t)$ can then be obtained by numerical integration of equations~\eqref{p:xi-Sec2}-\eqref{eq_sym_sec2-7}, respectively.
The results for three different parameter sets are shown in \figref{figure_Phi_xi_FT}: we find that the symmetry Eq.~\eqref{eq_FT_PDF_eff} holds over several orders of magnitude.
As previously discussed in this paper, the Gaussian approximation for the current $J_Y$ holds as long as the force $f_y$ is small. Given that our proof of the fluctuation relation Eq.~\eqref{eq_FT_PDF_eff} relies on this approximation, we expect a deviation from such a relation  as we increase the force. This is indeed what we observe by inspecting the panels in \figref{figure_Phi_xi_FT} from the leftmost to the rightmost one: for large values of $\xi$ there is an increasing discrepancy from the behavior predicted by Eq.~\eqref{eq_FT_PDF_eff}.

\section{$N$ coupled oscillators}
\label{Sec3}
We extend the model for a two particle machine described in~\secref{Sec2} and consider a system made up of $N$ overdamped Brownian particles coupled  through a periodic potential $U_0(x_1,\dots x_N)= \sum_{i,j} k_{i,j} u_0(x_i-x_j)$.
The dynamic equation for the $i$-th particle  reads
\begin{align}
\dot x_i=f_i-\sum_j k_{i,j}\partial_{x_i} u_0(x_i-x_j)+\zeta_i(t).
\label{eq_Saka}
\end{align}
 We assume uncorrelated Gaussian white noises, $\langle\zeta_i(t)\,\zeta_j(t')\rangle=2 T\delta_{ij}\delta(t-t')$, $i,j=1,\dots,N$. 
The case in which $k_{ij}=k,\,\forall i,j$, and $u_0(x)=-\cos(x)$ was  first introduced by Sakaguchi~\cite{Sakaguchi88} as an extension of the Kuramoto model~\cite{Kuramoto75,Kuramoto}. However, in the following we will not make any assumption on the specific form of the potential $U_0$.
As in the previous section, \secref{Sec2}, depending on the force sign, each oscillator can be considered either an energy producer ($f_i>0$) or an energy user ($f_i<0$). The single trajectory efficiency Eq.~\eqref{eq_eff-Intro} of this isothermal engine is then the rate between the work  extracted  by the users ($u$) and the work injected by the producers ($p$) along a single trajectory,
\begin{align}
\eta=-\frac{\sum_j^u f_jX_j}{\sum_i^p f_iX_i},
\label{eq_eff_NHO}
\end{align}
where we retain the notation as in the previous sections, and the capital letters indicate the unbounded coordinates. 
The superscripts appearing in the sum at the numerator and denominator of Eq.~\eqref{eq_eff_NHO} indicate that the sum is restricted to the users or producers, respectively.
Accordingly, the PDF of the efficiency reads
\begin{align}
P(\eta,t)=\int\D X_1\,\cdots\,\D X_N\,\delta\left(\eta+\frac{\sum_j^u f_jX_j}{\sum_i^p f_iX_i}\right)\,P(X_1,\dots,X_N,t),
\label{eq_PDF_eff_NHO}
\end{align}
where the PDF $P(X_1,\dots,X_N,t)$ depends implicitly on the forces $f_1,\dots f_N$.
Analogously to the unidimensional case Eq.~\eqref{eq_func_gen_lim_t}, we introduce the multidimensional version of the generating function 
\begin{align}
\psi({\underline\lambda},t)=\int \D X_1\,\cdots\,\D X_N\,\E^{\lambda_iX_i}\,P(X_1,\dots,X_N,t),
\label{eq_gen_func_multi-Sec3}
\end{align}
where underlined symbols represent vectors, and Einstein convention for the summation of repeated indexes is adopted. The generating function is dominated by the largest eigenvalue $\mu_0({\underline\lambda})$ of the multidimensional FP operator corresponding to the single coordinate operator Eq.~\eqref{eq_op_gen_func-Sec1-1},
\begin{align}
\psi({\underline\lambda},t)\sim\E^{t\mu_0({\underline\lambda})}.
\label{eq_psi-Sec3}
\end{align}
The fluctuation relation for the multidimensional PDF $P(X_1,\dots,X_N,t)$ implies the symmetry  for the largest eigenvalue~\cite{Fogedby12},
\begin{align}
\mu_0({\underline\lambda})=\mu_0(\{-\lambda_i-\beta f_i\}),\,~\forall \underline\lambda.
\label{eq_sym_FT_NHO}
\end{align}
In the following subsections we will explore the statistical properties of the efficiency for two different choices of the constant forces applied to the particles.

\subsection{Two terminals}
\label{Sec3-1}
In the first case that we consider  the system has two terminals, one where energy is injected and the other where it is extracted. We consider $f_1>0$ the input force and $f_N<0$ the output force, with $f_1>-f_N$, $f_i=0,~i\neq 1,N$. Then, the system of Langevin equations~\eqref{eq_Saka} reads
\begin{eqnarray}
\dot x_i&=&(\delta_{i,1}+\delta_{i,N}) f_i -\sum_j k_{i,j}\partial_{x_i} u_0(x_i-x_j)+\zeta_i(t). \label{eq_Langevin_2term_1-Sec3-1}
\end{eqnarray}
Thus, the stochastic efficiency Eq.~\eqref{eq_eff_NHO} reduces to
\begin{align}
\eta=-\frac{f_NX_N}{f_1X_1};
\end{align}
we introduce the rescaled efficiency
\begin{equation}
 \hat\eta\equiv-\frac{f_1}{f_N}\eta=\frac{X_N}{X_1},
\label{eq_eff-Sec3-1}
\end{equation}
whose PDF reads
\begin{eqnarray}
P(\hat\eta)&=&\int\int \D X_1\,\D X_N\,\delta\left(\hat\eta-\frac{X_N}{X_1}\right)\,\bar P(X_1,X_N,t),\label{eq_PDF_eff_N}\\
\bar P(X_1,X_N,t)&=&\int \D X_2\cdots \D X_{N-1}\,P(X_1,X_2,\cdots,X_{N-1},X_N,t),\nonumber
\end{eqnarray} 
and $P(X_1,X_2,\cdots,X_{N-1},X_N,t)$ is the solution of the FP equation associated to the Langevin equations~\eqref{eq_Langevin_2term_1-Sec3-1}.\\

In the long time limit, the dominant term of $\bar P(X_1,X_N,t)$ can be obtained trough saddle--point integration of the generating function,
\begin{align}
P(\underline X,t)\sim \int \D \underline{\lambda}\,\E^{\lambda_i X_i}\,\E^{t\mu_0(\underline{\lambda})}.
\end{align}
Hence
\begin{align}
\bar P(X_1,X_N,t)\propto\E^{t(\bar\mu_0(\lambda_1^*,\lambda_N^*)-J_1\lambda_1^*-J_N\lambda_N^*)},\label{eq_PDF_z1_zN}
\end{align}
with the saddle points implicitly defined by 
\begin{align}
\partial_{\lambda_1}\bar\mu_0(\lambda_1,\lambda_N)\Big|_{\lambda_1^*,\lambda_N^*}=J_1,\label{eq_lambda1_star}\\
\partial_{\lambda_N}\bar\mu_0(\lambda_1,\lambda_N)\Big|_{\lambda_1^*,\lambda_N^*}=J_N,\label{eq_lambdaN_star}
\end{align}
and where $J_1\equiv X_1/t\,,J_N\equiv X_N/t$, and $\bar\mu_0(\lambda_1,\lambda_N)$ is the cumulant generating function of $\bar P(X_1,X_N,t)$,
\begin{align}
\bar\mu_0(\lambda_1,\lambda_N)=\mu_0(\lambda_1,0,\dots,0,\lambda_N).
\label{eq_mu0_bar}
\end{align}
The details of the calculations are given in \appref{A3}.
 Plugging Eq.~\eqref{eq_PDF_z1_zN} into Eq.~\eqref{eq_PDF_eff_N}, and rearranging the terms in the integral, the PDF for the rescaled efficiency reads
\begin{align}
P(\hat\eta)\sim\int\int \D X_1\,\D X_N\,\delta\left(X_N-X_1\hat\eta\right)\,|X_1|\,\E^{t(\bar\mu_0(\lambda_1^*,\lambda_N^*)-J_1\lambda_1^*-J_N\lambda_N^*)}.
\label{eq_P_eff-Sec3-1}
\end{align}

\subsubsection{Extremal points}
\label{Sec3-1-1}
The extremal points of the efficiency's PDF 
correspond to the most likely efficiency,
\begin{align}
\eta_+=-\frac{f_N\langle J_N\rangle}{f_1\langle J_1\rangle},
\label{eq_most_likely_eff}
\end{align}
and the least likely efficiency,
\begin{align}
\eta_-=1,
\label{eq_least_likely_eff}
\end{align}
 see \appref{A3}.

Thus we conclude  that the study of the extremal points for the PDF of the efficiency of this machine made of $N$ all-to-all interacting oscillators with two terminals of input and output energy leads to the same efficiency features as for the two coupled oscillators machine studied in~\secref{Sec2-4}, namely the most likely efficiency is the macroscopic efficiency, whereas the least likely corresponds to the efficiency of the machine performing reversibly.


\subsubsection{Gaussian assumption}
In order to obtain an expression for the PDF of the efficiency Eq.~\eqref{eq_P_eff-Sec3-1} we need to assume a certain distribution for the variables $X_1$ and $X_N$. We assume thus a Gaussian distribution, that according to Eq.~\eqref{m0gaus1} implies the following expression for the cumulant generating function,
\begin{align}
\bar\mu_0(\lambda_1,\lambda_N)=\bar v_1\lambda_1(1+\lambda_1 T/f_1)+\bar v_N\lambda_N(1+\lambda_N T/f_N),
\label{eq_mu0_Gauss}
\end{align}
so as to fulfill the fluctuation relation Eq.~\eqref{eq_sym_FT_NHO}, and where $\bar v_i=\langle\dot X_i\rangle,\,(i=1,N)$.
Hence, $P(\eta,\tau)$ reads
\begin{align}
P(\eta,\tau)=\frac{\E^{-\tau/4}}{\pi a(\eta)\sqrt{|C|}}\{1+\sqrt{\pi\tau}\,h(\eta)\,\E^{\tau h^2(\eta)}\text{erf}(\sqrt{\tau}\,h(\eta))\}\,,
\label{eq_P_eta_Sec3}
\end{align}
with
\begin{eqnarray}
\tau&=&t\,\frac{f_N\overline{v}_N +f_1\overline{v}_1}{T},\nonumber \\
a(\eta)&=&(1-\eta)^2+\frac{1}{|C|}\left(\frac{\eta-\bar{\eta}}{1-\bar{\eta}}\right)^2,\nonumber \\
h(\eta)&=&\frac{1-\eta}{2\sqrt{a(\eta)}},\nonumber \\
C&=&\frac{1}{f_N\overline{v}_N+f_1\overline{v}_1}
\begin{bmatrix}
f_N\overline{v}_N  & 0\\
0 & f_1\overline{v}_1
\end{bmatrix},
\label{eq_P_eta_funcs_Sec3}
\end{eqnarray} 
and $\bar{\eta}=-f_N\bar v_N/(f_1\bar v_1)$.
The expression for $P(\eta,\tau)$ is the analogous of the one obtained in~\cite{Polettini15} for two coupled currents in the linear regime.

\begin{figure}[h]
\center
\psfrag{x}[ct][ct][1.]{$\eta$}
\psfrag{y}[cc][cc][1.]{$P(\eta,t)$}
\includegraphics[width=8cm]{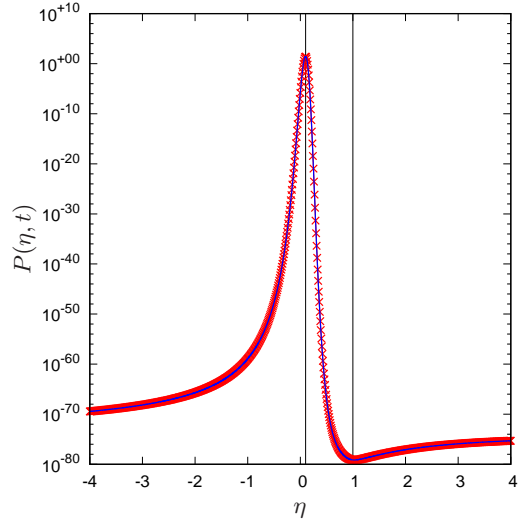}
\caption{PDF of the efficiency for a machine made of $N$ coupled oscillators with two terminals, whose currents are Gaussian distributed, and its fluctuation  symmetry. Red symbols: $P(\eta,\tau)$ Eq.~\eqref{eq_P_eta_Sec3}. Blue line: transformation of $P(\eta,\tau)$ according to the fluctuation relation Eq.~\eqref{FT_eta_2terminals}. Vertical black lines: extremal points of $P(\eta,\tau)$ Eqs.~\eqref{eq_most_likely_eff},\eqref{eq_least_likely_eff}. Parameters choice: $T=1,f_1=1.0,f_N=-0.2,\bar v_1=0.8,\bar v_N=0.4,\tau=1000$.}
\label{figure_P_eta_FT_2terminals}
\end{figure}

Retracing the steps in~\secref{Sec2-7}, we exploit the fluctuation symmetry for $\mu_0(\underline \lambda)$ Eq.~\eqref{eq_sym_FT_NHO} and find that the transformations
\begin{align}
g(\eta)=-\frac{\eta+(\eta-2)\bar\eta}{1-2\eta+\bar\eta},\label{eq_sym_map_2term}\\
\alpha(\eta) =\pm\left|\frac{1-2\eta+\bar\eta}{-1+\bar\eta}\right|,\label{eq_alpha_map_2term}
\end{align}
give the fluctuation relation
\begin{align}
P(\eta,\tau)=  P(g(\eta),\tau)\E^{-R(\eta)},
\label{FT_eta_2terminals}
\end{align}
with $R(\eta)=\ln \alpha^2(\eta)$.
This relation is graphically checked in \figref{figure_P_eta_FT_2terminals}, where we  also show the extremal points Eqs.~\eqref{eq_most_likely_eff}--\eqref{eq_least_likely_eff}, for a particular choice of the parameters.

The fluctuation relation for the efficiency has been derived  for an isothermal motor, with two energy currents coupled through a general potential. However, our results remain valid for other types of systems, for example the heat engine considered in \cite{Verley2014}, as long as the energy currents obey a fluctuation relation of the same type as Eq.~\eqref{eq_sym_FT_NHO}, where the term $\beta f_i$ is replaced by the corresponding generalized thermodynamic force associated with the current $J_i$.

\subsection{Distribution of forces}
\label{Sec3-2}
As a second case we consider a machine in which every oscillator is subject to a biasing force, so that we have a certain quenched distribution of input  ($f_i>0$) and output forces ($f_i<0$). The system of Langevin equations is the same as in Eq.~\eqref{eq_Saka}. Analogously to what we did in ~\secref{Sec2-3} for the relative coordinate, we assume a $N$-dimensional Gaussian PDF for $P(X_1,\dots,X_N,t)$, the cumulant generating function reads
\begin{align}
\mu_0({\underline\lambda})=\lambda_i\bar v_i+\alpha_{ij}\lambda_i\lambda_j,
\label{eq_SCGF-Sec3-2}
\end{align}
where $\bar v_i\equiv\langle \dot x_i\rangle$, and $\alpha_{ij}=\delta_{ij}\bar v_i T/f_i$ due to the symmetry imposed by the fluctuation relation Eq.~\eqref{eq_sym_FT_NHO}. As discussed in  ~\secref{Sec2-3} we expect such a Gaussian approximation to  hold in the limit of small forces. 

We define the input and output stochastic work as 
$W_{\text{out}}=\sum\nolimits_j^u f_jX_j$, $W_{\text{in}}=\sum\nolimits_i^p f_iX_i$.

Accordingly, the PDF of the efficiency Eq.~\eqref{eq_PDF_eff_NHO} reads
\begin{align}
P(\eta,t)=\int\int \D W_{\text{out}}\,\D W_{\text{in}}\,\delta\left(\eta+\frac{W_{\text{out}}}{W_{\text{in}}}\right)\,P(W_{\text{out}},W_{\text{in}},t).\label{eq_PDF_eff_NHOx}
\end{align}
Given Eq.\eqref{eq_SCGF-Sec3-2} one can easily check that the joint PDF on the right hand side of the last equation factorizes,  $P(W_{\text{out}},W_{\text{in}},t)=P(W_{\text{out}},t) P(W_{\text{in}},t)$ with 
\begin{align}
P(W_{\text{out(in)}},t)=\int\D\lambda \E^{t\lambda\left[P_{\text{out(in)}}-(1-\lambda T)\bar P_{\text{out(in)}}\right]},
\label{eq_Pworks-Sec3-2}
\end{align}
and where $\bar P_{\text{out(in)}}=\sum_j^{u(p)} f_j\bar v_j$ are the output (input) power, averaged over the force distribution. 
One thus obtains the bidimensional Gaussian distribution
\begin{align}
P(W_{\text{out}},W_{\text{in}},t)=\frac{1}{4\pi t T}\left(|\overline{P}_{\text{out}}\,\overline{P}_{\text{in}}|\right)^{-1/2}\exp\left[{-\frac{t}{4T}\left(\frac{(P_{\text{out}}-\overline{P}_{\text{out}})^2}{\overline{P}_{\text{out}}}+\frac{(P_{\text{in}}-\overline{P}_{\text{in}})^2}{\overline{P}_{\text{in}}}\right)}\right].
\end{align}
Therefore, the efficiency PDF Eq.~\eqref{eq_PDF_eff_NHOx} will be analogous to Eqs.\eqref{eq_P_eta_Sec3},\eqref{eq_P_eta_funcs_Sec3},
with
\begin{eqnarray}
\tau&=&\frac{t(\overline{P}_{\text{out}} +\overline{P}_{\text{in}})}{T},\nonumber \\
C&=&\frac{1}{\overline{P}_{\text{out}}+\overline{P}_{\text{in}}}
\begin{bmatrix}
\overline{P}_{\text{out}}  & 0\\
0 & \overline{P}_{\text{in}}
\end{bmatrix},
\end{eqnarray} 
and  $\bar\eta=-\overline{P}_{\text{out}}/\overline{P}_{\text{in}}$.
Accordingly, the fluctuation relation for the PDF of the efficiency is given by Eq.~\eqref{FT_eta_2terminals}, with the transformation Eqs.~\eqref{eq_sym_map_2term} and~\eqref{eq_alpha_map_2term}.

The statistical features of the efficiency of an isothermal engine made up of $N$-coupled oscillators, that are either producers or users according to a given distribution, are thus analogous to the statistical features of the efficiency in a device that couples two thermodynamic currents that fluctuate with normal law~\cite{Polettini15}. However, differently from ~\cite{Polettini15}, our system is not linear, the features of the non-linear interacting potential being hidden in the average velocities $\bar v_i$ that appear in Eq.~\eqref{eq_SCGF-Sec3-2}.


We end up this section by studying the extremal points of the PDF $\eta_{\pm}$.
They can be obtained by requiring that the transformation Eq.~\eqref{eq_sym_map_2term} maps each of them into itself, such that
\begin{align}
g(\eta_{\pm})=\eta_{\pm},
\end{align}
with $\alpha(\eta_{\pm})=1$.
According to this condition, the extremal points of the PDF Eq.~\eqref{eq_P_eta_Sec3} are $\eta_+=\bar\eta$ and $\eta_-=1$,
which correspond again to the macroscopic efficiency and the reversible efficiency, respectively.

However, we do not obtain the second maximum in the super-Carnot efficiency region  $\eta\ge \eta_-$ obtained in~\cite{Polettini15} in the intermediate time regime.  This is due to the fact that the long-time limit is already implicit in the derivation of the PDF $P(W_{\text{out}},W_{\text{in}},t)$ Eq.~\eqref{eq_P_eta_Sec3}.

\section{Conclusions}
\label{Conclusions}
We have studied the statistics of the efficiency in isothermal cyclic machines with realistic interactions between the internal degrees of freedom.
Such a realistic potential interaction has the advantage that we can consider explicitly the weak and the tight coupling limits as well as the small and large force limits.

We first investigate a minimal model consisting of two coupled degrees of freedom. By separating the center of mass and the relative coordinate motion, we are able to express the PDF of the efficiency as an integral of a closed form. Besides, we derive an analytic solution for the efficiency PDF in the limit of weak coupling and small forces.

The study of the extremal points of the efficiency PDF reveals that the most likely efficiency is always the macroscopic efficiency, whereas the least likely is the reversible efficiency.
The macroscopic efficiency, which depends on the interaction strength, is bounded between a minimal value obtained for weak coupling or strong forces, and a maximal value achieved in the tight coupling limit.
These boundaries turn out to be universal in the sense that they depend only on the thermodynamic forces, and not on the details of the interaction potential.

We investigate the condition under which the machine operates close to the macroscopic reversible efficiency, and we conclude that the tight coupling limit between the input and output currents is a necessary, yet not sufficient, condition for achieving the lossless limit. As a matter of fact, given the realistic physical interaction between the thermodynamic forces, the reversible macroscopic efficiency is attained in the tight coupling limit and close to the stall condition, in which the difference between the input and the output forces vanishes, thus making the machine useless.

Assuming a normal distribution for the relative coordinate current, the long time fluctuation relation for the input and output currents implies a fluctuation relation for the efficiency, that resembles the long time relations previously obtained for other stochastic thermodynamic quantities.
Even though this relation is derived under the conjecture of Gaussian distributed currents, whose range of validity is limited to the range of small forces and weak coupling, we provide numerical evidence that it holds for a wide range of forces, and hence beyond the linear regime.

We finally explore the case where the machine consists of $N$ degrees of freedom, and show that the efficiency fluctuations can be studied by focusing on the input and the output energy currents alone, i.e. mapping the $N$ body model into a model with two coupled fluctuating currents. Thus we find that the results obtained for the minimal model hold true for an arbitrary number of degrees of freedom.

\begin{acknowledgments}
We gratefully acknowledge the financial support of the Danish Council for Independent Research and of the Villum Foundation.
\end{acknowledgments}

\appendix
\section{Determinant near identity}
\label{A1}
The expansion of the second determinant in Eq.~\eqref{eq_recast-Sec1-2} is 
\begin{align}
\det\left[\left(\mathbb{1}+k\,{\bf M}^{-1}\,\hat {\bf L}_{\lambda}^{(1)}\right)\right]=\left[1+k\,f_1({\bf M}^{-1}\,\hat {\bf L}_{\lambda}^{(1)})+k^2\,f_2({\bf M}^{-1}\,\hat {\bf L}_{\lambda}^{(1)})+k^3\,f_3({\bf M}^{-1}\,\hat {\bf L}_{\lambda}^{(1)})+k^4\,f_4({\bf M}^{-1}\,\hat {\bf L}_{\lambda}^{(1)})+\mathcal{O}(k^5)\right],
\end{align}
 where the series expansion terms of the determinant near identity can be derived from the Jacobi's formula~\cite{Bellmann70,MagnusNeudecker07}, and by setting ${\bf A}\equiv {\bf M}^{-1}\,\hat {\bf L}_{\lambda}^{(1)}$ we have
\begin{eqnarray}
f_1({\bf A})&=&\text{Tr}[{\bf A}],\nonumber\\
f_2({\bf A})&=&\frac{\text{Tr}^2[{\bf A}]-\text{Tr}[{\bf A}^2]}{2}, \nonumber\\
f_3({\bf A})&=&\frac{1}{3!}(\text{Tr}^3({\bf A})-3\text{Tr}({\bf A})\text{Tr}({\bf A}^2)+2\text{Tr}({\bf A}^3)), \nonumber\\
f_4({\bf A})&=&\frac{1}{4!}\text{Tr}^4({\bf A})-\frac{1}{4}\text{Tr}({\bf A}^4)+\frac{1}{8}\text{Tr}^2({\bf A}^2)+\frac{1}{3}\text{Tr}({\bf A})\text{Tr}({\bf A}^3)-\frac{1}{4}\text{Tr}^2({\bf A})\text{Tr}({\bf A}^2).\nonumber
\end{eqnarray}
The matrix ${\bf M}^{-1}$ depends on $\mu(\lambda)$ too, and thus the terms $f_i({\bf M}^{-1}\,\hat {\bf L}_{\lambda}^{(1)})$ have to be expanded in powers of $k$ as well,  so as to take into account 
all the contributions for each in $k$.

\section{PDF of the efficiency for the Gaussian approximation}
\label{A2}
We assume a normal distribution for the relative coordinate $Y$, $P(Y,t)$ Eq.~\eqref{pygaus1-Sec2-3}.  After integrating  Eq.~\eqref{p:xi-Sec2}, the PDF of $\xi$ reads 
\begin{align}
\Phi(\xi,t)=\exp\left[-t\left(\frac{f_x^2+f_y \bar v_y}{2T}\right)\right]\frac{\sqrt{f_y \bar v_y}}{\pi (f_y+\bar v_y\xi^2)}\left(1 +\E^{t \hat h(\xi)^2} \sqrt{\pi t}\, \hat h(\xi) \erf{\left(\sqrt{t}\, \hat h(\xi)\right)}\right),
\label{eq_phi_xi-AppB}
\end{align}
where $\hat h(\xi)=(f_y+f_x\xi)\left(2 T(f_y+\bar v_y\xi^2)/\bar v_y\right)^{-1/2}$. In the long time limit the error function can be expanded $\erf(\sqrt{t}\, \hat h(\xi))\sim 1-\E^{-t\,\hat h(\xi)^2}/(\sqrt{\pi t}|\hat h(\xi)|)$~\cite{AbramowitzStegun65}; taking into account that the $\erf{(\sqrt{t}\,\hat h(\xi))}$ change sign at $\hat h(\xi)=0$, the long time limit of $\Phi(\xi,t)$ Eq.~\eqref{eq_phi_xi-AppB} reads
\begin{align}
\Phi(\xi,t)=\exp\left[{-t\left(\frac{(f_x^2+f_y \bar v_y)}{2T}\right)}\right] \frac{\sqrt{f_y \bar v_y}}{\pi(f_y+\bar v_y\xi^2)}\left(1-\frac{\hat h(\xi)}{| \hat h(\xi)|}+\sqrt{\pi t}|\hat h(\xi)|\E^{t\,\hat h(\xi)^2}\right).
\label{eq_phi_xi_t-AppB}
\end{align}
We obtain the PDF of $\eta$ after inverting the change of variables Eq.~\eqref{eq_xi-Sec2},
\begin{align}
P(\eta,t)=\exp\left[{-t\left(\frac{(f_x^2+f_y \bar v_y)}{2T}\right)}\right]\frac{4 f_y T\, h(\eta)^2}{(f_x+f_y)\pi\sqrt{f_y \bar v_y}(\eta-1)^2|f_x-f_y|}\left(1+\sqrt{\pi t}\,h(\eta)\,\E^{t\, h(\eta)^2}\erf{\left(\sqrt{t}\, h(\eta)\right)}\right),
\label{eq_P_eta-AppB}
\end{align}
where $h(\eta)=(f_x^2-f_y^2)(\eta-1)\sqrt{\bar v_y}\left((f_y(\eta-1)+f_x(\eta+1))\sqrt{2T\left(f_y+\frac{\bar v_y(f_x(\eta-1)+f_y(\eta+1))^2}{(f_y(\eta-1)+f_x(\eta+1))^2}\right)}\right)^{-1}$. Applying the former expansion for the error function, the long time limit PDF of the efficiency reads
\begin{align}
P(\eta,t)=\exp\left[{-t\left(\frac{(f_x^2+f_y \bar v_y)}{2T}\right)}\right]\frac{4 f_y T\, h(\eta)^2}{(f_x+f_y)\pi\sqrt{f_y \bar v_y}(\eta-1)^2|f_x-f_y|}\left(1-\frac{h(\eta)}{|h(\eta)|}+\sqrt{\pi t}\,|h(\eta)|\,\E^{t\,h(\eta)^2}\right).
\end{align}


\section{PDF of the efficiency for two terminals and its extremal points}
\label{A3}
The leading term of the integral in Eq.~\eqref{eq_P_eff-Sec3-1} is
\begin{align}
\G(\hat\eta)=\int\int\int \D s\,dX_1\,dX_N\,e^{t[\mu_0(\lambda_1^*,\lambda_N^*)-J_1\lambda_1^*-J_N\lambda_N^*+\ii s(J_N-J_1\hat\eta)]}\,,\label{eq_leading_PDF}
\end{align}
where we have used the integral expression for the Dirac delta,
\begin{align}
\delta\left(X_N-X_1\hat\eta\right)=\frac{1}{2\pi}\int \D s\, e^{\ii s t(J_N-J_1\hat\eta)}.
\label{eq_Dirac_delta-Sec3-1}
\end{align}
Integrating over $X_1$ and $X_N$ by the saddle-point approximation we obtain
\begin{align}
\G(\hat\eta)=\int d s \,\E^{t[\mu_0(\lambda_1^{**},\lambda_N^{**})-J_1^{**}\lambda_1^{**}-J_N^{**}\lambda_N^{**}+\ii s(J_N^{**}-J_1^{**}\hat\eta)]},\label{eq_leading_PDF_2}
\end{align}
with $J_1^{**}$ and $J_1^{**}$ implicitly defined by the equations
\begin{align}
\partial_{J_1}\left[\mu_0(\lambda_1^*,\lambda_N^*)-\lambda_1^*J_1-\lambda_N^*J_N+\ii s(J_N-J_1\hat\eta)\right]\Big|_{J_1^{**},J_N^{**}}=0,\\
\partial_{J_N}\left[\mu_0(\lambda_1^*,\lambda_N^*)-\lambda_1^*J_1-\lambda_N^*J_N+\ii s(J_N-J_1\hat\eta)\right]\Big|_{J_1^{**},J_N^{**}}=0.
\end{align}
By employing the conditions that define $\lambda_1^*$ Eq.~\eqref{eq_lambda1_star} and $\lambda_N^*$ Eq.~\eqref{eq_lambdaN_star}, and labeling $\lambda_1^{**}\equiv\lambda_1^{*}(J_1^{**},J_N^{**})$ and $\lambda_N^{**}\equiv\lambda_N^{*}(J_1^{**},J_N^{**})$, the equations that define $J_1^{**}$ and $J_N^{**}$ are
\begin{align}
-\lambda_1^{**}-\ii s \hat\eta=0\,,\label{eq_z1_star}\\
-\lambda_N^{**}+\ii s =0\,.\label{eq_zN_star}
\end{align}
The integral over $s$ in Eq.~\eqref{eq_leading_PDF_2} can  be solved by a saddle--point approximation as well
\begin{align}
\G(\hat\eta)=\E^{t[\mu_0(\lambda_1^{***},\lambda_N^{***})-J_1^{***}\lambda_1^{***}-J_N^{***}\lambda_N^{***}+\ii  s^{***}(J_N^{***}-J_1^{***}\hat\eta)]},\label{eq_g_Sec-3-1}\\
\partial_{s}\left[\mu_0(\lambda_1^{**},\lambda_N^{**})-\lambda_1^{**}J_1^{**}-\lambda_N^{**}J_N^{**}+\ii s(J_N^{**}-J_1^{**}\hat\eta)\right]\Big|_{s^{***}}=0.
\end{align}
Taking into account Eqs.~\eqref{eq_lambda1_star},~\eqref{eq_lambdaN_star},~\eqref{eq_z1_star}, and~\eqref{eq_zN_star}, the condition for $s^{***}$ can be rewritten after some algebraic manipulation as 
\begin{align}
J_N^{***}-J_1^{***}\hat\eta=0\,.\label{eq_xi_star}
\end{align}

The extremal points of the efficiency's PDF will be given by those of $\G(\hat\eta)$ Eq.~\eqref{eq_g_Sec-3-1}, that is, the solution of
\begin{align}
\partial_{\hat\eta}\left[\mu_0(\lambda_1^{***},\lambda_N^{***})-\lambda_1^{***}J_1^{***}-\lambda_N^{***}J_N^{***}+\ii s^{***}(J_N^{***}-J_1^{***}\hat\eta)\right]\Big|_{\hat\eta^{***}}=0\,,
\end{align}
that simplifies into
\begin{align}
s^{***}J_1^{***}=0,\label{eq_eta_star}
\end{align}
after applying Eqs.~\eqref{eq_lambda1_star},~\eqref{eq_lambdaN_star},~\eqref{eq_z1_star},~\eqref{eq_zN_star}, and~\eqref{eq_xi_star}. The two solutions of Eq.~\eqref{eq_eta_star} are
\begin{align}
s^{***}=0\,,~J_1^{***}=0.
\end{align}
When $s^{***}=0$, then $\lambda_1^{***}=\lambda_N^{***}=0$ Eqs.~\eqref{eq_z1_star},~\eqref{eq_zN_star}.
Plugging Eq.~\eqref{eq_psi-Sec3} into Eq.~\eqref{eq_gen_func_multi-Sec3}, deriving with respect to $\lambda_i$ and evaluating at $\lambda_1=\lambda_N=0$, we obtain the identity (analogous to Eq.~\eqref{eq_vss_mu0-Sec1-3})
\begin{align}
\partial_{\lambda_i}\mu_0(\underline\lambda)\Big|_{\lambda_1=\lambda_N=0}=\frac{\langle X_i\rangle}{t}=\langle J_i\rangle.
\label{eq_vss_mu0_multi}
\end{align}
Exploiting Eq.~\eqref{eq_vss_mu0_multi},
we can compute $J_1^{***}$ and $J_N^{***}$ appearing in Eq.~\eqref{eq_xi_star} from Eqs.~\eqref{eq_lambda1_star},~\eqref{eq_lambdaN_star}. Then we can solve Eq.~\eqref{eq_xi_star} for $\hat\eta$ and we find that the most likely efficiency is,
\begin{align}
\hat\eta_+=\frac{\langle J_N\rangle}{\langle J_1\rangle}\Rightarrow \eta_+=\frac{-f_N\langle J_N\rangle}{f_1\langle J_1\rangle},
\label{eq_most_likely_eff_App}
\end{align}
where the transformation in Eq.~\eqref{eq_eff-Sec3-1} has been taken into account.\\
Considering the second solution $J_1^{***}=0$, then $J_N^{***}=0$ because of Eq.~\eqref{eq_xi_star}. Bearing in mind that the largest eigenvalue is a convex function, then $(\lambda_1^{***},\lambda_N^{***})$ are the coordinates of its minimum, for $J_i^{***}=0$ implies that $\partial_{\lambda_i}\mu_0(\lambda_1,\lambda_N)\Big|_{\lambda_1^{***},\lambda_N^{***}}=0$ according to Eqs.~\eqref{eq_lambda1_star},~\eqref{eq_lambdaN_star}. Then the symmetry imposed by the fluctuation relation Eq.~\eqref{eq_sym_FT_NHO} is such that the symmetry point, i.e. the minimum, is located at $(-f_1/2T,-f_N/2T)$. Thus the least likely efficiency is,
\begin{align}
\eta_-=1\,;
\label{eq_least_likely_eff_App}
\end{align}
where the definitions of $J_1^{**}$ Eq.~\eqref{eq_z1_star} and $J_N^{**}$ Eq.~\eqref{eq_zN_star} have been employed, together with Eq.~\eqref{eq_eff-Sec3-1}.\\

\bibliography{bibliography}

\end{document}